\documentclass{siamart1116}
\usepackage{amsfonts}
\usepackage{graphicx}
\usepackage{epstopdf}
\usepackage{algorithmic}
\ifpdf
  \DeclareGraphicsExtensions{.eps,.pdf,.png,.jpg}
\else
  \DeclareGraphicsExtensions{.eps}
\fi

\numberwithin{theorem}{section}

\usepackage{etoolbox}
\AtBeginEnvironment{bmatrix}{\setlength{\arraycolsep}{1pt}}

\usepackage[utf8]{inputenc}
\usepackage{amsmath,amsfonts,amssymb}
\usepackage{mathrsfs}  
\usepackage{graphicx}  
\usepackage{booktabs}
\usepackage{adjustbox}
\usepackage{pgfplots}
\usepackage{subfig}
\pgfplotsset{compat=1.5}

\newcommand{\review}[1]{\textcolor{black}{{#1}}}

\newcommand\scalemath[2]{\scalebox{#1}{\mbox{\ensuremath{\displaystyle #2}}}}

\newcommand*{\bm}[1]{\boldsymbol{#1}}
\newcommand{\aled}[1]{\widehat{#1}}
\newcommand{\aleop}{X}

\allowdisplaybreaks

\makeatletter
\renewcommand*\env@matrix[1][*\c@MaxMatrixCols c]{%
  \hskip -\arraycolsep
  \let\@ifnextchar\new@ifnextchar
  \array{#1}}
\makeatother

\newcommand{\TheTitle}{A monolithic ALE Newton-Krylov solver 
    with Multigrid-Richardson-Schwarz preconditioning
     for incompressible Fluid Structure Interaction} 
\newcommand{\TheAuthors}{E. Aulisa, S. Bn\`a, and G. Bornia}

\headers{A monolithic ALE Newton-Krylov solver for incompressible FSI}{\TheAuthors}

\title{{\TheTitle}\thanks{Submitted to the editors Feb 11, 2017.
\funding{This work was partly supported by the National Science Foundation grant DMS-1412796.}}}

\author{
  Eugenio Aulisa\thanks{Department of Mathematics and Statistics, Texas Tech University, Lubbock, TX
    (\email{eugenio.aulisa@ttu.edu})}
  \and
  Simone Bn\`a\thanks{CINECA - SCAI (Super Computing Applications and Innovation), Casalecchio di Reno, BO, Italy (\email{simone.bna@cineca.it})}
  \and
  Giorgio Bornia\thanks{Department of Mathematics and Statistics, Texas Tech University, Lubbock, TX
    (\email{giorgio.bornia@ttu.edu})}
}

\begin{document}

\maketitle

 \begin{abstract}
In this paper we study a monolithic Newton-Krylov solver 
with exact Jacobian for the solution of incompressible FSI problems.
A main focus of this work is on the use of geometric multigrid preconditioners
with modified Richardson smoothers
preconditioned by an additive Schwarz algorithm. 
The definition of the subdomains in the Schwarz smoother is driven 
by the natural splitting between fluid and solid.
 The monolithic approach
 guarantees the automatic satisfaction of the stress balance and the kinematic conditions 
across the fluid-solid interface. 
The enforcement of the incompressibility conditions both for the fluid and for the solid parts 
is taken care of by using inf-sup stable finite element pairs without stabilization terms.
A suitable Arbitrary Lagrangian Eulerian (ALE) operator 
is chosen in order to avoid mesh entanglement
while solving for large displacements of the moving fluid domain. 
Numerical results of two and three-dimensional benchmark tests 
with Newtonian fluids and nonlinear hyperelastic solids
show a robust performance of our fully incompressible solver 
especially for the more challenging direct-to-steady-state problems.
\end{abstract}



\section{Introduction}

 Fluid-Structure Interaction (FSI) problems
 are of paramount interest because of a number of ubiquitous applications. 
 \review{To give an idea of such a breadth, 
 we recall examples from aeroelasticity \cite{liu2001calculation,tezduyar2008fluid,bazilevs20113d}, 
hydroelasticity \cite{young2008fluid,LeTallec,Tijsseling2007844},
biomechanics \cite{HronTurek, formaggia2010cardiovascular}, 
civil engineering \cite{szabo2009three,koh1998fluid,wiggert2001fluid},
acoustics \cite{everstine1990coupled}, 
poroelasticity \cite{detournay1989poroelasticity}. 
Several research groups at international level 
have dedicated their efforts to the study of fluid-structure interactions
in universities, research institutes as well as industries.
From this interest 
many conferences and workshops
have been organized in the last decades,
 journals have been established and books published \cite{bungartz2006fluid,bungartz2010fluid, formaggia2010cardiovascular}.
Furthermore, software projects of both open-source and commercial type
have been developed in order to perform numerical simulations of FSI phenomena \cite{bathe2006finite}.
}

Given a certain physical model for the solid and fluid parts,
many challenging questions are still nowadays open in the FSI community, 
ranging from experimental investigations \cite{gomes2011experimental,benchexp2015}
to theoretical analysis \cite{BeiraodaVeiga,Grandmont,kukavica2012solutions},
numerical approximation and computational issues.
Fluid-structure interaction problems are characterized by an intrinsic mathematical challenge, 
due to the inherent nonlinearity
given by a domain that moves as a function of the unknowns.
Several choices are possible in terms of
the identification of the fluid and solid moving domains (interface tracking or capturing),
the definition of the coupling algorithm between fluid and solid
(monolithic vs. partitioned, loosely coupled vs. strongly coupled),
the discretization procedure (decouple-then-discretize or \review{vice versa}),
the order between the discretization and the linearization procedures (linearize-then-discretize or \review{vice versa}),
the nonlinear loop (fixed-point, relaxed fixed-point, quasi-Newton, Newton), 
the choice of the linear solvers and preconditioners.

In this work we focus on the performance of geometric multigrid preconditioners combined with domain decomposition smoothers
for monolithic Newton-Krylov solvers of FSI saddle-point problems of either steady-state or time-dependent type. 
We intend to highlight the effectiveness of our algorithms
in handling two kinds of numerical difficulties concerning FSI simulations: 
the enforcement of incompressibility (both for the fluid and for the solid part), 
and the computation of direct-to-steady-state solutions.
We enforce pure incompressibility conditions both for the fluid and for the solid using inf-sup stable finite element pairs, 
 without introducing slightly-compressible stabilization terms.
 \review{To the best of our knowledge, this is the first contribution
 in the literature on the study of this class of preconditioners 
 for the case of incompressibility both in the fluid and in the solid part.}
Concerning steady-state solutions, our algorithms are able to perform direct-to-steady-state computations. 
This is an advantage with respect to the more time-consuming practice
of using pseudo-time stepping schemes, 
in which the linear systems to be solved have a better conditioning 
and a stationary solution is reached as a limit of a time sequence.

Both multigrid and domain decomposition methods draw a lot of attention within the FSI community.
Multigrid algorithms are taken into account for the solution of large sparse linear systems 
due to their optimal computational complexity, which can be proven rigorously for model elliptic problems \cite{Brenner}. 
Domain decomposition methods are very appealing since they allow for an effective parallel implementation,
in which several local subproblems can be solved over subdomain patches. 
In \cite{HronTurek} a geometric multigrid solver
with a Multilevel Pressure Schur Complement (MPSC) Vanka-like smoother is considered,
with applications to nonstationary FSI problems in biomechanics.
Applications of this scheme to hemodynamics are also addressed in \cite{TurekBenchmark2,razzaq2012fem}.
Monolithic Newton-Krylov algorithms are studied in \cite{gee2011truly,wu2013parallel}.
In \cite{gee2011truly} the inner Krylov iterations are preconditioned with algebraic multigrid methods,
while an overlapping additive Schwarz preconditioner 
is considered in \cite{wu2013parallel} with application to parallel three-dimensional blood flow simulations.
A partitioned method in which multigrid is used either within the fluid and solid solvers 
or as an outer iteration is addressed in \cite{schafer2006implicit,sternel2008efficiency}.
A work that is closest to ours and that may be seen as a starting point for our contribution is \cite{Richter}. 
Following this work, 
we also solve the coupled problem in a monolithic manner at each level
and we perform partitioning between fluid and solid only at the smoothing level within the multigrid preconditioner.
The idea of this approach is to
invert smaller matrices with better condition numbers in the smoothing process \cite{Richter}.
However, in \cite{Richter} the smoothing is partitioned 
but without using domain decomposition algorithms within the solid and fluid domains.
Also, differently from our work, \cite{Richter} makes use of a pressure stabilization terms
and the ALE equation is simply taken as a harmonic operator.
Ultimately, \cite{Richter} only deals with time-dependent problems. 
Other numerical studies are available in the literature on the use of domain decomposition 
Vanka-type smoothers for multigrid both in Computational Fluid Dynamics (CFD) 
and in Computational Solid Mechanics (CSM) 
\cite{AMS,TUREK,VANKAOrig,wobker2009numerical,razzaq2012fem,AulisaDD2}. 

In order to deal with the several nonlinearities inherent to FSI problems
(advection terms,
 transformations between moving and fixed domains, 
 nonlinear constitutive relations),
in this work we use an exact Newton method where we compute the exact Jacobian matrix 
with automatic differentiation tools provided by the Adept software package \cite{hoganadept}.
 For the purposes of the implementation, automatic differentiation 
 is a very convenient tool that can be exploited with little code modification.
 Analytic expressions of the exact Jacobian may also be implemented
  using shape derivative calculus \cite{fernandez2005newton,richter2012goal}.
 In certain cases it may be more convenient, for simplicity or time performance,
 to consider the use of approximate Jacobians.
In \cite{HronTurek} quasi-Newton outer iterations with line search are performed
and the Jacobian matrix is computed by a divided difference approach.
A quasi-Newton method in which the variation 
of the fluid domain in the fluid equations 
is neglected is proposed in \cite{SimoneICNAAM,SimonePhD}.
In \cite{Gerbeau} the authors propose a quasi-Newton algorithm based 
on a reduced model for fluid-structure interaction problems. 
 
For the movement of the solid and fluid domains,
we describe the solid motion in a Lagrangian way, while the fluid is observed in Eulerian fashion.
We use the ALE approach,  
which is one of the most popular techniques in the FSI community \cite{Donea,sackinger1996newton,wick2011fluid,fernandez2009derivation}
and it differs from other approaches 
such as the immersed boundary method \cite{Peskin2} or the fully Eulerian approach \cite{dunne_fullyeul}.
A judicious definition of the ALE operator is needed in certain conditions 
in order to preserve the mesh quality.
In this work we follow an approach from \cite{tezduyar1994_mesh_update}
to define a convenient linear elastic operator
In our numerical experiments we observed that this approach is very robust at preserving the orientation of the mesh elements, 
or in other words it avoids mesh entanglement (see also \cite{SimonePhD}).
 
The paper is organized as follows. 
In Section \ref{form} we present the strong and weak formulations of
 the time-dependent and stationary incompressible FSI problems under investigation.
We describe the linearization procedure by means of automatic differentiation,
and we illustrate the features of the multigrid preconditioner with domain decomposition smoothing.
Numerical results of benchmark problems are presented in Section \ref{sec_numres}.
Finally, we draw our conclusions.

\section{Formulation of the incompressible FSI problem}
\label{form}

Here we describe the mathematical formulation of the FSI problem. 
We first define deformation mappings and displacement fields. 
Then, we describe the fluid-structure interaction problem
in terms of three subproblems with mutual coupling.
For more details, 
we refer the reader to \cite{fernandez2009derivation,HronTurek,fernandez2005newton,boffi2004stability}.

\subsection{Deformation mappings and displacement fields}

For every time $ t \in [0,T]$,
let $\Omega^f_t \subset \mathbb{R}^n $ be an open set occupied only by a fluid,
and let $\Omega^s_t \subset \mathbb{R}^n $ be an open set occupied only by a solid.
In the following, any other symbol endowed with the superscripts $ f $ or $ s $ 
will refer to either the fluid or the solid part, respectively.
We denote the boundary of the fluid and solid parts 
as $ \partial \Omega^f_t $ and  $ \partial \Omega^s_t $, respectively. 
We define the parts of the boundary adjacent only to the fluid or only to the solid as $\Gamma^f_t$ and $\Gamma^s_t$,
such that $ \partial \Omega^f_t = \Gamma^f_t \cup \Gamma^i_t $ 
and $ \partial \Omega^s_t = \Gamma^s_t \cup \Gamma^i_t $.
The symbols $ \bm{n}^f $ and $ \bm{n}^s $ denote the outward unit normal fields defined on $ \partial \Omega^f_t$ and $ \partial \Omega^s_t $.

We now define the open set $\Omega_t := \Omega^f_t \cup \Omega^s_t \cup \Gamma^i_t $, 
which is the current configuration of the overall physical domain,
where $ \Gamma^i_t $ is the \textit{interface} between fluid and solid.
The fluid and solid are immiscible, namely 
$ {\Omega^f_t} \cap {\Omega^s_t} = \emptyset \,, $ 
and they interact through the nonempty interface $ \Gamma^i_t = \partial {\Omega^f_t} \cap \partial {\Omega^s_t} $.
For every domain $ D_t \subset \mathbb{R}^n $ (which may change in time), we also define the cylinder
$    Q_D = \{ (\bm{x},t) \text{ s.t. } \bm{x} \in D_t, t \in [0,T] \} \,. $

We use the \textit{hat} notation to define $ \widehat{\Omega}^f := \Omega^f_0 $ and $ \widehat{\Omega}^s := \Omega^s_0 $.
Normally, they are referred to as the \textit{undeformed} or \textit{stress-free} configurations,
although the initial stresses
 need not be identically zero either in the solid or in the fluid part.
Moreover, we define $ \widehat{\Omega} := \Omega_0$ and $ \widehat{\Gamma}^i:= \Gamma^i_0 $.
The domains $ \{ \widehat{\Omega}^s, Q_{\widehat{\Omega}^s} \} $ are called \textit{Lagrangian domains}
and the fields $ \widehat{q}^s(\widehat{\bm{x}}) $ or $ \widehat{q}^s(\widehat{\bm{x}}, t) $ defined on them 
are called \textit{Lagrangian fields}. 
The domain $ \widehat{\Omega}^s $ is initially occupied by the solid we observe.
We follow the motion of the solid in a Lagrangian way.
The domains $ \{ \widehat{\Omega}^f , Q_{\aled{\Omega}^f} \} $ are called \textit{ALE domains}
and the fields $ \aled{q}^f(\aled{\bm{x}}) $ or $ \aled{q}^f({\bm{x}}, t)$ defined on them 
are called \textit{ALE fields}.
The domain $ \widehat{\Omega}^f $ is the domain on which we initially observe the fluid motion in a Eulerian way.

\review{
As a consequence of the solid movement, the domain on which we observe the fluid motion 
changes in time as well, so that we need to define a deformation for the fluid domain.
The domain $ \Omega^f_t $ is occupied only by fluid at each time $ t $. 
The moving fluid or solid domains $ {\Omega^f_t} $ and $ {\Omega^s_t} $ and the corresponding cylinders are called \textit{Eulerian domains},
and fields  $ q({\bm{x}}) $ or $ q({\bm{x}}, t)$ defined on Eulerian domains are called \textit{Eulerian fields}.  
For the sake of brevity, when no confusion arises, we will denote fields 
on the fixed domains (Lagrangian or ALE) with the hat notation $ \widehat{q} $
and Eulerian fields as $ q $ with no symbol on top,
without specifying space and time arguments.
In the same spirit, we will use the notations $\widehat{\nabla}$ or $\nabla$ to refer to
the nabla symbolic operator in the fixed or moving frames, respectively.
}

%


\begin{figure}
\centering
 \includegraphics[width=0.8\textwidth]{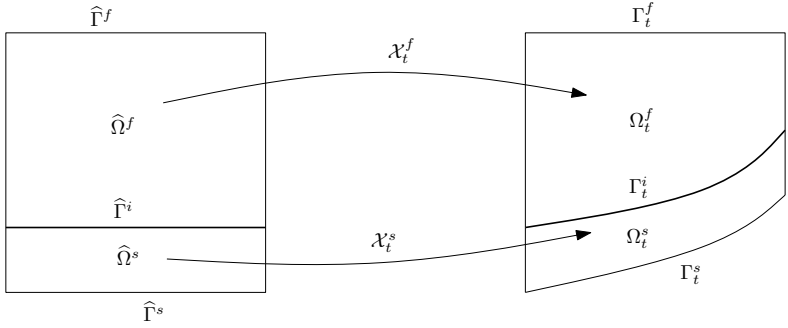}
 \caption{Mappings between fixed (Lagrangian or ALE) and moving (Eulerian) domains}
\end{figure}

\review{
In order to describe the motion of the fluid and solid domains, 
we define a $ t $-parametrized family of invertible and sufficiently regular mappings $ \mathcal{X}_t $,
called \textit{deformation mappings}, given by a perturbation of the identity, so that
\begin{align}
 \mathcal{X}_t & : \widehat{\Omega} \rightarrow \Omega_t \,,  \quad \mathcal{X}_t(\widehat{\bm{x}})  := \widehat{\bm{x}} + \widehat{\bm{d}}(\widehat{\bm{x}},t)  \,.
\end{align}
The field $ \widehat{\bm{d}}(\widehat{\bm{x}},t) $ is called \textit{displacement field}.
 For every $ (\widehat{\bm{x}},t) \in \, Q_{\widehat{\Omega}} $, we also define
\begin{align}
&  \widehat{\bm{F}} ( \widehat{\bm{d}}(\widehat{\bm{x}},t) )  = \widehat{\nabla}\mathcal{X}_t(\widehat{\bm{x}}) = I + \widehat{\nabla} \widehat{\bm{d}}(\widehat{\bm{x}},t) \,, \\ 
& \widehat{J}      ( \widehat{\bm{d}}(\widehat{\bm{x}},t) )  = \det \widehat{\bm{F}}( \widehat{\bm{d}}(\widehat{\bm{x}},t) ) \,,
 \quad \widehat{\bm{B}} ( \widehat{\bm{d}}(\widehat{\bm{x}},t) )  = \widehat{\bm{F}} ( \widehat{\bm{d}}(\widehat{\bm{x}},t) ) \widehat{\bm{F}}^T ( \widehat{\bm{d}}(\widehat{\bm{x}},t) ) \,.
\end{align}
The symbols  $ \widehat{\bm{F}} $ and  $ \widehat{\bm{B}} $ denote 
the \textit{deformation gradient tensor}
and the \textit{left Cauchy-Green deformation tensor}, respectively.
}

\review{
For a given physical quantity $ q $,  the deformation mapping and its inverse 
allow one to move between the fixed (Lagrangian or ALE) descriptions 
and the moving (Eulerian) descriptions of a given physical quantity $ q $, namely,
\begin{align}
 q({\bm{x}},t)  = \widehat{q}(\mathcal{X}_t^{-1}(\bm{x}),t) \,,\quad
 \widehat{q}(\widehat{\bm{x}},t)  = {q}(\mathcal{X}_t(\widehat{\bm{x}}),t)  \,.
\end{align}
}




\review{
The displacement field $\widehat{\bm{d}}(\widehat{\bm{x}},t)$ is determined separately in the fluid and solid parts 
as a solution of two different subproblems.
Its restrictions $  \aled{\bm{d}}^f(\aled{\bm{x}},t) $ and $ \widehat{\bm{d}}^s(\widehat{\bm{x}},t) $ 
are referred to as \textit{fluid domain displacement} (or \textit{ALE displacement})
and \textit{solid displacement}, respectively.
They are required to take on common values at the interface, namely
\begin{align}
  \widehat{\bm{d}}^s(\widehat{\bm{x}},t) = \aled{\bm{d}}^f(\aled{\bm{x}},t)   \,,\quad  \widehat{\bm{x}} \in \widehat{\Gamma}^i
  \,.
\end{align}
For the solid, $\widehat{\bm{d}}^s(\widehat{\bm{x}},t)$ is determined by the solution of the \textit{solid subproblem} 
consisting of the elasticity equations.
Therefore, $ \widehat{\bm{d}}^s(\widehat{\bm{x}},t) $
is the actual \textit{material} displacement at time $t$ of a solid particle 
that was originally in the fixed position $ \widehat{\bm{x}} $.
On the other hand,
  $\aled{\bm{d}}^f(\aled{\bm{x}},t) $ is an \textit{artificial} displacement of the fluid domain
with no physical meaning, that is used to map the deformation 
of the fluid domain. 
}

Some additional definitions are in order. 
We define the \textit{fluid domain velocity} $ \aled{\bm{w}}^f(\aled{\bm{x}},t) $  as
\begin{align}
 \aled{\bm{w}}^f(\aled{\bm{x}},t)   =  \frac{\partial \mathcal{\aleop}^f_t(\aled{\bm{x}})}{\partial t} =  
 \frac{\partial \aled{\bm{d}}^f(\aled{\bm{x}},t) }{\partial t} \,, \quad  (\aled{\bm{x}},t) \in \, Q_{\aled{\Omega}^f} \,.
\end{align}

Moreover, the \textit{ALE time derivative} of every Eulerian field $ q(\bm{x},t) $ 
is denoted as $\frac{\partial q}{\partial t} |_{\mathcal{\aleop}^f}$ 
and is given for every $  (\bm{x},t) \in Q_{\Omega_t^f} $ by \cite{fernandez2009derivation} 
\begin{equation}
\frac{\partial q(\bm{x},t)}{\partial t}\biggl|_{\mathcal{\aleop}^f} = \frac{d }{dt}\left( q(\mathcal{\aleop}^f_t(\aled{\bm{x}}),t) \right) = 
\frac{\partial q(\bm{x},t)}{\partial t} + ( \bm{w}^f(\bm{x},t) \cdot \nabla)q(\bm{x},t)
\,,\quad   \bm{x}=\mathcal{\aleop}^f_t(\aled{\bm{x}})\,,\; 
 \end{equation}
where $\frac{\partial q(\bm{x},t)}{\partial t}$ is the usual \textit{Eulerian time derivative}
and $ \bm{w}^f(\bm{x},t)$ is the Eulerian representation of the fluid domain velocity,
\begin{align}
  {\bm{w}}^f({\bm{x}},t)   = \aled{\bm{w}}^f( (\mathcal{\aleop}^f_t)^{-1}({\bm{x}}),t) \,, \quad  ({\bm{x}},t) \in \, Q_{\Omega^f_t} \,.
\end{align}

\subsection{Strong FSI problem} 
\label{sec_fsistrong}

A Fluid-Structure Interaction problem can be formulated in terms of three subproblems:
the \textit{fluid subproblem}, 
the \textit{solid subproblem} and 
the \textit{subproblem for the fluid domain displacement}.
Each of them possesses  \textit{interface}, \textit{boundary} and \textit{initial} conditions.

In a monolithic formulation,
we define three unknown fields (displacement, velocity and pressure)
in a piecewise fashion
at each point of the Eulerian cylinder $ Q_{\Omega_t}$ as
\begin{align}
& \bm{d} =
\begin{cases}
  {\bm{d}}^s  \text{ in } Q_{\Omega^s_t} \\
  {\bm{d}}^f  \text{ in } Q_{\Omega^f_t} \,,  
\end{cases} 
\bm{u}  =
\begin{cases}
{\bm{u}}^s  \text{ in } Q_{\Omega^s_t} \\
{\bm{u}}^f  \text{ in } Q_{\Omega^f_t} \,,
\end{cases} 
  p  =
\begin{cases}
{p}^s  \text{ in } Q_{\Omega^s_t}  \\
  p^f  \text{ in } Q_{\Omega^f_t}  \,.
\end{cases} 
\end{align}

The fields 
$ {\bm{d}}^s $, $ {\bm{d}}^f $,
$ {\bm{u}}^s $, $ {\bm{u}}^f $,
 $ {p}^s $, $ {p}^f $
are involved in the solution of the three subproblems,
which will be described in the following.
In a monolithic formulation the fields $ {\bm{d}} $, $ {\bm{u}} $ and $ p $
are computed in an implicit way. 
\review{Monolithic algorithms are the most robust and stable 
among the strong coupling approaches (\cite{Heil,HronTurek}).}
As will be pointed out in the definition of the subproblems, $ \bm{d} $ and  $ \bm{u} $ 
take on the same values on the fluid-solid interface,
while $ p $ has no continuity conditions across the interface.


The \textit{solid subproblem} consists in determining $ (\widehat{\bm{d}}^s(\widehat{\bm{x}},t), \widehat{p}^s(\widehat{\bm{x}},t)) $
as solutions of
\begin{align}
 & \widehat{\rho}^s \widehat{J}( \widehat{\bm{d}}^s ) \frac{\partial^2 \widehat{\bm{d}}^s}{\partial t^2}
   -  \widehat{\nabla} \cdot \left( \widehat{J}( \widehat{\bm{d}}^s ) \widehat{\bm{\sigma}}^s(\widehat{\bm{d}}^s,\widehat{p}^s)   (\widehat{\bm{F}}(\widehat{\bm{d}}^s))^{-T} \right) - \widehat{\rho}^s \widehat{J}( \widehat{\bm{d}}^s ) \widehat{\bm{f}}^s = \bm{0} \quad  \text{ in } \, Q_{\widehat{\Omega}^s} \,,  \\  
 & \widehat{J}( \widehat{\bm{d}}^s ) - 1  = 0    \quad  \text{ in } \, {Q}_{\widehat{\Omega}^s} \,, \\
 & \widehat{J}( \widehat{\bm{d}}^s ) \widehat{\bm{\sigma}}^s ( \widehat{\bm{d}}^s,\widehat{p}^s ) (\widehat{\bm{F}}(\widehat{\bm{d}}^s))^{-T} \cdot \widehat{\bm{n}}^s = - \bm{\sigma}^f \left( {\bm{u}}^f,{p}^f \right)  \cdot \bm{n}^f \quad   \text{ on } Q_{\widehat{\Gamma}^i} \,,  \\
 & \mathcal{B}^s_t(\widehat{\bm{d}}^s(\widehat{\bm{x}},t),\widehat{p}^s(\widehat{\bm{x}},t)) = \bm{0}    \quad  \text{ on }  Q_{\widehat{\Gamma}^s}\,, \\
 & \widehat{\bm{d}}^s(\widehat{\bm{x}},0) = \bm{0}  \quad   \text{ in } \Omega^s_0 \,,  \\
 & \frac{\partial \widehat{\bm{d}}^s }{ \partial t}    (\widehat{\bm{x}},0) = \bm{0}  \quad  \text{ in } \Omega^s_0 \,.
\end{align}
The first two equations are the solid momentum and mass balances written in Lagrangian form, 
also known as the incompressible elasticity equations.  
The symbols $\rho^s$ and $\bm{f}^s$  denote mass density and body force density for the solid, respectively.
At the interface with the fluid, we enforce a Neumann condition of continuity of the normal stress.
For the sake of generality, 
we denoted with $\mathcal{B}^s_t$ an abstract boundary operator for the solid boundary $ \widehat{\Gamma}^s$,
which may correspond to Dirichlet, Neumann or other types of boundary conditions.
We observe that the initial displacement is zero by definition, and we consider a zero initial material velocity.

 For the solid stress tensor  $ \widehat{\bm{\sigma}}^s $ we consider hyperelastic models for large strains,
either incompressible Neo-Hookean or incompressible Mooney-Rivlin, 
whose Lagrangian description is given for every $ (\widehat{\bm{x}},t) \in Q_{\widehat{\Omega}^s} $ by
\begin{align}
\widehat{\bm{\sigma}}^s_{NH} ( \widehat{\bm{d}}^s , \widehat{p}^s ) 
   & = - \widehat{p}^s \bm{I} + 2 C_1 \widehat{\bm{B}}( \widehat{\bm{d}}^s ) \;,
\\
\widehat{\bm{\sigma}}^s_{MR} (\widehat{\bm{d}}^s , \widehat{p}^s )
   & = - \widehat{p}^s  \bm{I} + 2 C_1 \widehat{\bm{B}}(\widehat{\bm{d}}^s ) -  2 C_2 ( \widehat{\bm{B}}(\widehat{\bm{d}}^s ) )^{-1} \;,
\end{align}
where the constants $C_1$ and $C_2$ depend on the mechanical properties of the material.
Clearly, the tensors are by definition symmetric.
We remark that the pressure in the solid $ \widehat{p}^s $ does not have a clear physical meaning 
and can be regarded mathematically 
as the Lagrange multiplier associated 
to the solid incompressibility constraint.
With the given choices of stress tensors, the momentum balance is of second-order in the space derivatives.
It is also of second order in the time derivative for the displacement unknown.
 In order to obtain a system of equations containing only first-order time derivatives,
  we also introduce the \textit{solid velocity} 
 $ \widehat{\bm{u}}^s(\widehat{\bm{x}},t)  $ 
 which is the velocity at time $ t $ of the solid particle initially at $ \widehat{\bm{x}} $,
 defined as
\begin{align} \label{strong_solid_vel}
  \widehat{\bm{u}}^s(\widehat{\bm{x}},t) 
  & = \frac{\partial \mathcal{X}^s_t(\widehat{\bm{x}})}{\partial t}
    = \frac{\partial \widehat{\bm{d}}^s(\widehat{\bm{x}},t)}{\partial t}  \,.   
\end{align}
Then, this last equation
 is enforced over the solid region, including the solid-fluid interface $ \widehat{\Gamma^i} $.
The stress at the interface is the input to the solid subproblem coming from the fluid part.
The output of this subproblem is the displacement of the solid $ \widehat{\bm{d}}^s $.
The pressure $ \widehat{p}^s $ is an internal variable.
Clearly, the overall fluid-solid coupling is two-way, 
as the solid displacement at the interface 
modifies the fluid domain, thus affecting the fluid velocity.


The unknowns  $ ({\bm{u}}^f({\bm{x}},t), p^f({\bm{x}},t)) $ of the \textit{fluid subproblem} are solutions of
\begin{align}
 & \rho^f \biggl( \frac{\partial \bm{u}^f}{\partial t} \biggl|_{\mathcal{\aleop}^f}  + [ ( \bm{u}^f - \frac{\partial \bm{d}^f}{\partial t} ) \cdot \nabla ] \bm{u}^f \biggl) - \nabla \cdot \bm{\sigma}^f({\bm{u}}^f,{p}^f)  - \rho^f  \bm{f}^f = \bm{0}   \quad    \text{ in } \, Q_{\Omega^f_t} \,, \\
 & \nabla \cdot \bm{u}^f = 0  \quad      \text{ in } \, Q_{\Omega^f_t} \,, \\
 &    \bm{u}^f({\bm{x}},t)  = \widehat{\bm{u}}^s( \mathcal{X}_t^{-1}( {\bm{x}} ),t) \quad        \text{ on }  Q_{\Gamma^i_t} \,,  \\
 & \mathcal{B}^f_t( {\bm{u}}^f({\bm{x}},t), {p}^f({\bm{x}},t)) = \bm{0}    \quad   \text{ on }  Q_{\Gamma^f_t}\,,  \\
 & \bm{u}^f({\bm{x}},0)  = \bm{u}_0({\bm{x}}) \quad   \text{ in }  \Omega^f_0\,.
\end{align}
The first two equations are the fluid momentum and mass balances,
referred to as the incompressible Navier-Stokes equations.
Here, $\rho^f$ and $\bm{f}^f$  are mass density and body force density for the fluid. 
We enforce the continuity of velocity at the solid-fluid interface.
We remark that the fluid subproblem is of Dirichlet type on the interface,
while the solid subproblem is of Neumann type.
In a sense, the formulation of a FSI problem 
 can be interpreted as a nonverlapping domain decomposition formulation of Dirichlet-Neumann type.
Similarly as before, we denoted with $\mathcal{B}^s_t$ an abstract boundary operator for the fluid boundary.
 The initial velocity profile is denoted as $\bm{u}_0({\bm{x}})$.
 
The fluid stress tensor $ \bm{\sigma}^f $ for incompressible Newtonian fluid flows is given 
as a Eulerian field for every $ ({\bm{x}},t) \in Q_{\Omega^f_t} $ by
\begin{align}  \label{newtonian}
\bm{\sigma}^f ( \bm{u}^f , p^f ) = - p^f \bm{I} + \mu ( \nabla \bm{u}^f + (\nabla \bm{u}^f)^T ) \,,
\end{align}
where $\mu$ is the fluid viscosity.
We remark that the momentum balance for the fluid is of first order in the time derivative for the velocity unknown.
With the given stress tensor, it is second-order in the space derivatives.

{Input} to the fluid subproblem is the displacement of the fluid domain.
 This is used to compute the position of each point as
\begin{equation}
  \bm{x} = \aled{\bm{x}} + \aled{\bm{d}}(\aled{\bm{x}},t).
\end{equation}
Moreover, the displacement of the fluid domain is used to compute the ALE time derivative and the fluid domain velocity.
Another input is given from the solid side by the solid displacement on the interface,
whose time derivative gives the Dirichlet condition on velocity.
 The {outputs} of this system are both fluid velocity and fluid pressure, 
 which are used to compute the stress at the interface for the solid subproblem.

The \textit{subproblem for the fluid domain displacement} consists in determining
the unknown $ \aled{\bm{d}}^f(\aled{\bm{x}},t) $ as a solution of
\begin{align} 
 & \aled{\nabla} \cdot \left( k(\aled{\bm{{x}}}) ( \aled{\nabla} \aled{\bm{d}}^f + ( \aled{\nabla} \aled{\bm{d}}^f )^T  ) \right) = \bm{0} \,, \quad  \text{ in }  \, Q_{\aled{\Omega}^f} \,,  \label{ale_eqn} \\
 &  \aled{\bm{d}}^f = \widehat{\bm{d}}^s \,, \quad   \text{ on } {Q}_{\aled{\Gamma}^i} \,, \\
 & \mathcal{B}^{fd}_t(\aled{\bm{d}}^f( \aled{\bm{x}},t))  = \bm{0} \,, \quad     \text{ on } {Q}_{\aled{\Gamma}^f}  \,.
\end{align}
This subproblem is also referred to as the \textit{kinematic equation} 
or the \textit{pseudo-solid mapping} \cite{sackinger1996newton}, 
as it defines the arbitrary motion of the fluid domain as another elastic solid.
At the solid-fluid interface we enforce a Dirichlet condition of continuity of the displacement.
The geometry of the fluid-solid interface is determined by the solution of the elasticity equations.
 We denote with $\mathcal{B}^{fd}_t$ a general boundary operator which can be chosen arbitrarily depending on the problem at hand.
 Dirichlet, Neumann or mixed Dirichlet-Neumann conditions are some of the possible choices.
We remark that we consider an equation without time derivative of the fluid displacement. Nevertheless, the solution depends on time
because of the time-dependent interface condition.
The boundary operator $\mathcal{B}^{fd}_t$ may also be chosen to depend on time.

\begin{figure}
  \centering
 \includegraphics[width=0.45\textwidth]{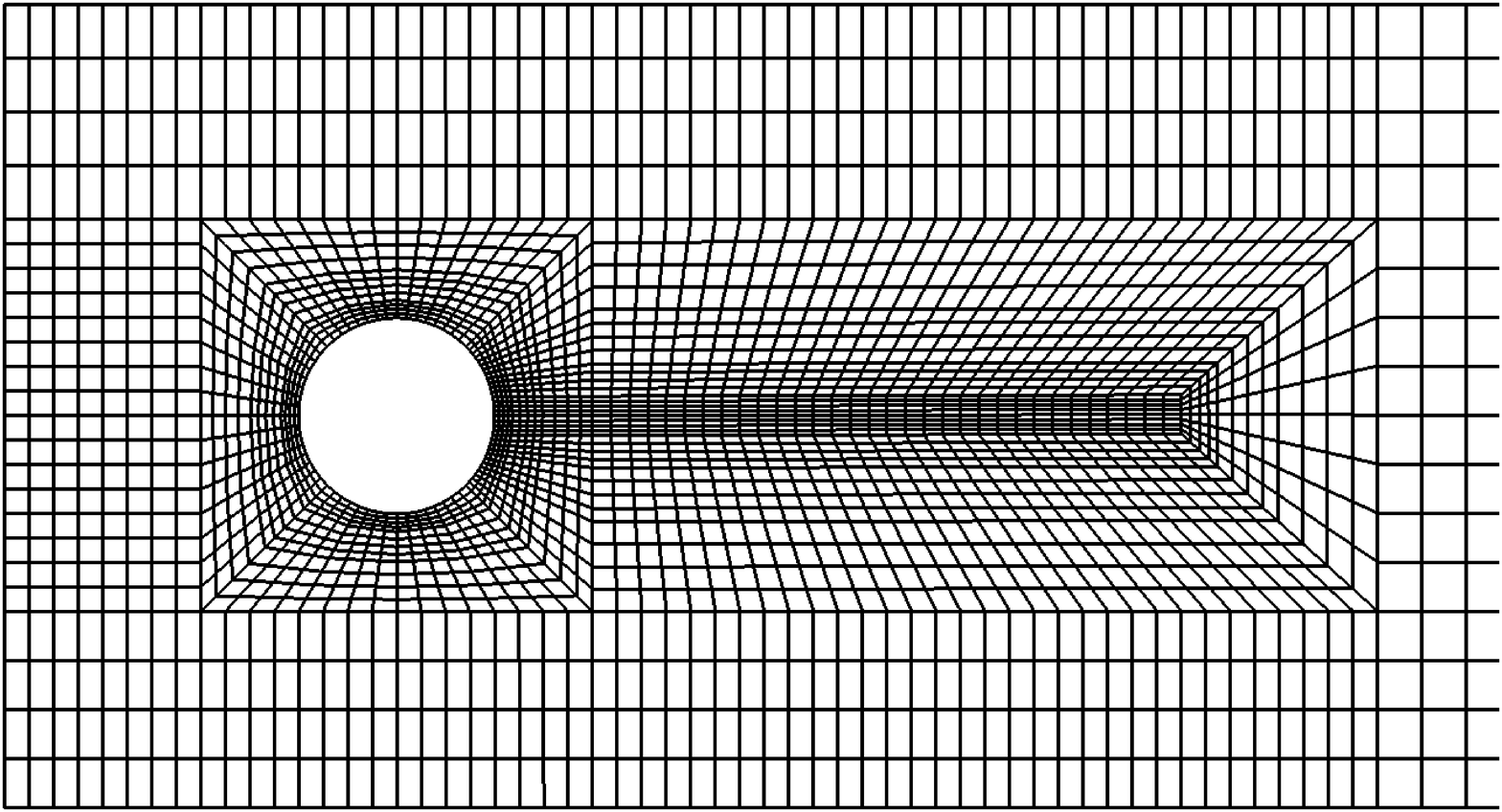}\hspace{0.25cm}
 \includegraphics[width=0.45\textwidth]{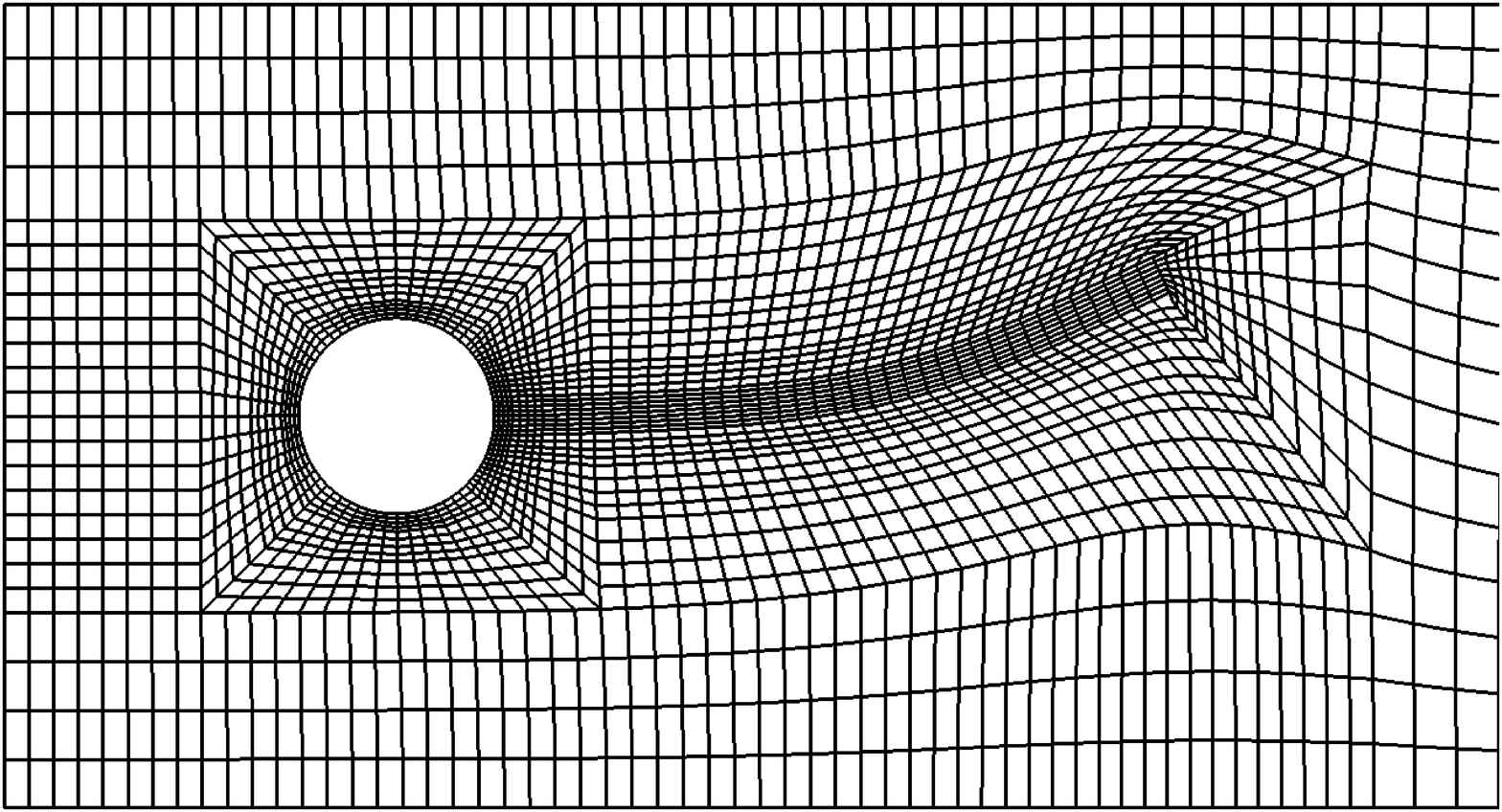}
  \caption{An example of mesh associated to the ALE operator \eqref{ale_eqn} for a typical FSI benchmark configuration} 
  \label{fine_mesh_ale}
\end{figure}

In this work the mesh deformation is based on the sizes of the elements.
Following \cite{tezduyar1994_mesh_update}, 
we want smaller elements to be stiffer than larger ones.
In regions where the mesh is expected to undergo large distortions (e.g. the region near the fluid-solid interface) 
we use smaller elements, in order not to degrade the mesh quality.
Smaller elements can be made stiffer by setting the function $ k(\bm{{x}}) $ 
to be a piecewise-constant function discontinuous across the element boundary and whose value is given by 
\begin{equation}
 k(\bm{{x}}) = \frac{1}{\mbox{V}_{el(\bm{{x}})}},
\end{equation}
where $\mbox{V}_{el}$ is the volume of the mesh element that contains the $\bm{{x}}$ coordinate.
An example of suitable mesh for this definition of the ALE operator \eqref{ale_eqn} is given in Figure \ref{fine_mesh_ale} 
for a classical FSI configuration of flow around a cylinder with deformable flap. 

This subproblem receives as a sole {input} the displacement of the fluid-solid interface from the solid part.
The displacement of the fluid domain is the {output}. 
This is used to update the ALE mapping $ \mathcal{\aleop}^f_t $,
with which the ALE domain velocity and the ALE time derivative are computed in the fluid subproblem. 
 
\subsection{Weak FSI problem}

Here we describe the weak formulation of the monolithic FSI problem.
For the sake of simplicity, we will denote with the same symbol $ ( \cdot, \cdot )  $ 
the standard inner products 
 either on $ L^2(\mathcal{O}) $,
 $ L^2(\mathcal{O})^n $
or $ L^2(\mathcal{O})^{n \times n} $,
for any open set $ \mathcal{O} \in \mathbb{R}^n$.
 On the fluid-only boundary $ {\Gamma}^f_t $, 
 we denote with $ {\Gamma}^f_{t,D} $ and $ {\Gamma}^f_{t,D,\bm{d}^f} $ the subsets of $ {\Gamma}^f_t $
 on which Dirichlet boundary conditions on the velocity $\bm{u}^f $ 
 and on the fluid domain displacement $\bm{d}^f $ are enforced, respectively.
 Similarly, 
  given the solid-only boundary $ {\Gamma}^s_t $ 
 we denote with $ {\Gamma}^s_{t,D} $ the part
 on which Dirichlet boundary conditions on the displacement $\bm{d}^s $ are enforced. 
In order to keep the exposition simple, 
we do not discuss the case of mixed Dirichlet-Neumann conditions
and the related definitions of function spaces and variational equations. 
Boundary conditions have been dealt with in a standard manner in this work
so that no significant feature has to be pointed out.
For any boundary subset $ \Gamma \subseteq  \partial\mathcal{O} $, we denote with $ {H}^1_0(\mathcal{O};\Gamma) $
 the subspace of functions in $ {H}^1(\mathcal{O}) $ with zero trace on $ \Gamma $.
 \review{We use the boldface notation to denote $ n $ copies of the corresponding non-boldface function spaces,
 where $n$ is the space dimension.}
 Now define 
 \begin{align}
 & \bm{V}^{\bm{f}}  :=  \bm{H}^1(\Omega^f_t) \,, \quad
   \bm{V}^{\bm{s}}  :=  \bm{H}^1(\Omega^s_t)
   \,, \\ 
 & \bm{V}^{\bm{f}}_0              :=  \bm{H}^1_{0}(\Omega^f_t;         {\Gamma}^f_{t,D} ) \;, 
   \bm{V}^{\bm{s}}_0              :=  \bm{H}^1_{0}(\Omega^s_t;         {\Gamma}^s_{t,D} ) \;, 
   {\bm{V}}^{\bm{f}}_{0,\bm{d}^f} :=  \bm{H}^1_{0}({\Omega}^f_t; {\Gamma}^f_{t,D,\bm{d}^f} )
   \,, \\ 
 & \bm{V} := 
  \{ \bm{v} = (\bm{v}^f,\bm{v}^s) \in 
  \bm{V}^{\bm{f}} \times \bm{V}^{\bm{s}} \text{ s. t. } \bm{v}^f = \bm{v}^s \text{ on } \Gamma^i_t \}  \,, \\
 & \bm{V}_0  := 
  \{ \bm{v} = (\bm{v}^f,\bm{v}^s) \in  
  \bm{V}^{\bm{f}}_0 \times \bm{V}^{\bm{s}}_0 \text{ s. t. } \bm{v}^f = \bm{v}^s \text{ on } \Gamma^i_t \} \,. 
 \end{align}
 The mapping of $ {\bm{V}}^{\bm{f}}_{0,\bm{d}^f} $ to the reference domain 
 is denoted as $\widehat{\bm{V}}^{\bm{f}}_{0,\bm{d}^f}$.
 In order to obtain a monolithic weak form of the momentum balance equations
 that automatically satisfies the interface stress balance, 
 let the momentum operator $ M(\bm{d}, \bm{u}, p) $ be given in a piecewise way by
\begin{equation}
  M(\bm{d}, \bm{u}, p) =
  \begin{cases}
    \rho^f \biggl( \dfrac{\partial \bm{u}}{\partial t} \biggl|_{\mathcal{\aleop}^f}  + \left[ \left( \bm{u} - \dfrac{\partial \bm{d}}{\partial t} \right) \cdot \nabla \right] \bm{u} \biggl) - \nabla \cdot \bm{\sigma}^f(\bm{u},p)  - \rho^f  \bm{f}^f   \; \text{ in } \, Q_{\Omega^f_t}, \\
    \widehat{\rho}^s \widehat{J}( \widehat{\bm{d}} ) \dfrac{\partial \widehat{\bm{u}}}{\partial t} -  \widehat{\nabla} \cdot ( \widehat{J}( \widehat{\bm{d}} ) \widehat{\bm{\sigma}}^s(\widehat{\bm{d}},\widehat{p}) (\widehat{\bm{F}}(\widehat{\bm{d}}))^{-T} ) - \widehat{\rho}^s \widehat{J}( \widehat{\bm{d}} ) \widehat{\bm{f}}^s   \quad   \text{ in } \, Q_{\widehat{\Omega}^s}\,. 
 \end{cases}
\end{equation}
 Then, the monolithic weak form of the momentum balance can be written as 
 \begin{equation} \label{weakmom}
  \left( M(\bm{d}, \bm{u}, p) ,  \bm{\phi}^{\bm{m}} \right)_{\Omega_t} = 0 \quad \forall  \bm{\phi}^{\bm{m}} \in \bm{V}_0 \,.
 \end{equation}

 The monolithic weak FSI problem consists in finding 
 $ (\bm{d}, \bm{u}, p) $ in 
 $ \bm{V} \times \bm{V} \times L^2(\Omega_t) $ 
 solution of a system that can be split into three parts:
 the \textit{weak momentum balance} given by \eqref{weakmom}, i.e.
\begin{multline}
  \left( \widehat{\rho}^s \widehat{J}( \widehat{\bm{d}} ) \frac{\partial \widehat{\bm{u}}}{\partial t} , \widehat{\bm{\phi}}^{\bm{m}}         \right)_{\widehat{\Omega}^s}
 + \left( \widehat{J}( \widehat{\bm{d}} ) \widehat{\bm{\sigma}}^s(\widehat{\bm{d}},\widehat{p}) (\widehat{\bm{F}}(\widehat{\bm{d}}))^{-T}            , \widehat{\nabla} \widehat{\bm{\phi}}^{\bm{m}}  \right)_{\widehat{\Omega}^s} 
 - \left( \widehat{\rho}^s \widehat{J}( \widehat{\bm{d}} ) \widehat{\bm{f}}^s    ,  \widehat{\bm{\phi}}^{\bm{m}}        \right)_{\widehat{\Omega}^s} \\
 + \frac{d}{dt} \left( \rho^f \bm{u} , \bm{\phi}^{\bm{m}} \right)_{\Omega^f_t } 
 -  \left( \rho^f  \left( \nabla \cdot \frac{\partial \bm{d}}{\partial t} \right)  \bm{u}             , \bm{\phi}^{\bm{m}} \right)_{\Omega^f_t}
 +  \left( \rho^f [ (\bm{u} - \frac{\partial \bm{d}}{\partial t} ) \cdot \nabla ] \bm{u} , \bm{\phi}^{\bm{m}} \right)_{\Omega^f_t}  \\
+  \left( \sigma^f (\bm{u},p)                                                    , \nabla \bm{\phi}^{\bm{m}} \right)_{\Omega^f_t}
-  \left( \rho^f \bm{f}^f                                                               , \bm{\phi}^{\bm{m}} \right)_{\Omega^f_t}
  = 0 
 \qquad
 \forall \; \bm{\phi}^{\bm{m}} \in \bm{V}_0 \,, 
\end{multline} 
the \textit{weak mass balance}
\begin{align}
\left( \widehat{J}( \widehat{\bm{d}} ) - 1 , \widehat{\phi}^{ps} \right)_{\widehat{\Omega}^s} 
 & = 0 
 \qquad
\forall \; \widehat{\phi}^{ps} \in L^2(\widehat{\Omega}^s),
\label{cntsld} \\ 
 \left( \nabla \cdot  \bm{u}  , \phi^{pf} \right)_{{\Omega}^f_t} 
 & = 0
\qquad
\forall \; \phi^{pf} \in L^2({\Omega}^f_t)
\label{cntfld}
\end{align}
and the \textit{weak kinematic equations}
\begin{align} 
 \left( \widehat{\bm{u}} - \frac{\partial \widehat{\bm{d}}}{\partial t}  , \widehat{\bm{\phi}}^{\bm{ks}} \right)_{\widehat{\Omega}^s} 
  & = 0
 \;,
 \forall \; \widehat{\bm{\phi}}^{\bm{ks}} \in \bm{H}^1(\widehat{\Omega}^{s}), 
 \label{kinsld} \\  
 \left( k(\aled{\bm{{x}}})  \left(  \aled{\nabla} \aled{\bm{d}}^f + (\aled{\nabla} \aled{\bm{d}}^f)^T \right), \aled{\nabla} \widehat{\bm{\phi}}^{\bm{kf}}  \right)_{\widehat{\Omega}^f}
  & = 0
\;,
\forall \; \widehat{\bm{\phi}}^{\bm{kf}} \in \bm{H}_0^1(\widehat{\Omega}^{f}; \widehat{\Gamma}^i) \cap \widehat{\bm{V}}^{\bm{f}}_{0,\bm{d}^f} \,.
\label{kinfld}
 \end{align}
 \review{Eq. \eqref{kinsld} is the weak form of the solid velocity equation \eqref{strong_solid_vel}.}
The {interface physical condition} of normal stress continuity 
is enforced in the monolithic weak momentum balance, where the boundary integrals disappear due to the condition
\begin{equation}
   \sigma^s (\bm{d},p) \bm{n}^s + \sigma^f (\bm{u},p) \bm{n}^f = 0 \quad \text{ on } {\Gamma}^i_t\,.
\end{equation}
Concerning velocity continuity, notice that the solid kinematic equation is just a change of variables 
without associated boundary conditions and its test functions are in $ \bm{H}^1(\widehat{\Omega}^{s}) $.
The interface velocity computed from this equation is an input for the fluid momentum balance.
With reference to displacement continuity,
notice also that the test functions in the weak fluid domain displacement equation are in 
$ \bm{H}_0^1(\widehat{\Omega}^{f}; \widehat{\Gamma}^i) \cap \widehat{\bm{V}}^{\bm{f}}_{0,\bm{d}^f} $,
so that they vanish on the solid-fluid interface.
Thus, this equation does not affect the value of the 
displacement on the interface, which is an unknown of the problem 
that is evaluated by solving the other parts of the system.

We remark that in deriving the weak FSI system we used the identity
\begin{equation}
 \left( \rho^f \frac{\partial \bm{u}}{\partial t}\biggl|_{\mathcal{\aleop}^f} , \bm{\phi} \right)_{\Omega^f_t } =
  \frac{d}{dt} \left( \rho^f \bm{u} , \bm{\phi} \right)_{\Omega^f_t } 
 -  \left( \rho^f  \left( \nabla \cdot \frac{\partial \bm{d}}{\partial t} \right)  \bm{u} , \bm{\phi} \right)_{\Omega^f_t} 
\end{equation}
that is proved in \cite{fernandez2009algorithms}, p. 312; see also \cite{boffi2004stability}.
The advantage of this formula is that one need not compute time derivatives along the ALE deformations.

\subsection{Operator FSI problem}

The weak monolithic FSI problem can be rewritten in a more compact form in terms 
of operators that are naturally associated to the fluid, solid and interface parts.
We describe this operator formulation
in order to later describe the matrix blocks obtained after discretization in terms of these operators.
We modify some ideas as in \cite{fernandez2009algorithms,fernandez2005newton}.
Unlike \cite{fernandez2009algorithms}, continuity of displacement and velocity at the interface
is enforced in a strong way, without weak boundary integrals on the interface.
This operator formulation is based on a characterization of the space $ \bm{V}_0 $.
First, given $ \widehat{\bm{g}} \in \bm{H}^{1/2} ({\Gamma}^i_t) $,
we define three $t$-parametrized families of operators $ \mathcal{L}^f_t $, $ \mathcal{L}^s_t $ and $ \mathcal{R}_t $ as
\begin{align*}
  \mathcal{L}^f_t  : \bm{H}^{1/2} ({\Gamma}^i_t) \rightarrow \bm{H}^{1}_0(D^f_t;\partial D^f_t \setminus {\Gamma}^i_t) \,,  & \quad
  \mathcal{L}^s_t  : \bm{H}^{1/2} ({\Gamma}^i_t) \rightarrow \bm{H}^{1}_0(D^s_t;\partial D^s_t \setminus {\Gamma}^i_t) \,,  \\
  \mathcal{R}_t    : \bm{H}^{1/2} ({\Gamma}^i_t) & \rightarrow \bm{H}_0^1(D^f_t \cup D^s_t) \,.
\end{align*}
Here,  $ \mathcal{L}^f_t $, $ \mathcal{L}^s_t $ are lifting operators, 
 while $ D^f_t \subset {\Omega}^f_t $ and $ D^s_t \subset {\Omega}^s_t $ are subsets whose boundaries contain $ {\Gamma}^i_t $.
The two liftings are put together by $ \mathcal{R}_t {\bm{g}} := (\mathcal{L}^f_t {\bm{g}}, \mathcal{L}^s_t {\bm{g}}) $.
%
We also define
\begin{align}
 & {\bm{W}}^s_0 :=   \bm{V}^{\bm{s}}_0 \cap \bm{H}^1_0({\Omega}^s_t;{\Gamma}^i_t) \,,\quad
 \widetilde{\bm{W}}^s_0  
  := \{ \bm{v} : \bm{v}|_{{\Omega}^s_t} \in {\bm{W}}^s_0, \bm{v} = \bm{0} \text{ on } \Omega_t \backslash \Omega^s_t \}   \,, \\
 & {\bm{W}}^f_0 := \bm{V}^{\bm{f}}_0  \cap \bm{H}^1_0({\Omega}^f_t;{\Gamma}^i_t)  \,,\quad
 \widetilde{\bm{W}}^f_0
  := \{ \bm{v} : \bm{v}|_{{\Omega}^f_t} \in {\bm{W}}^f_0, \bm{v} = \bm{0} \text{ on } \Omega_t \backslash \Omega^f_t \} \,, \\
 & {\bm{W}}^f_{0,\bm{d}^f} :=  \bm{V}^{\bm{f}}_{0,\bm{d}^f} \cap \bm{H}^1_0({\Omega}^f_t;{\Gamma}^i_t) \,, \\  
 & {\bm{W}} 
  := \{ (\mathcal{L}^f_t {\bm{g}},\mathcal{L}^s_t {\bm{g}} )\,, {\bm{g}} \in  \bm{H}^{1/2} ({\Gamma}^i_t) \} \,.
\end{align}
We have the direct sum decomposition (see \cite{fernandez2009algorithms}, p. 330) 
\begin{equation}
 {\bm{V}}_0 = 
 \widetilde{\bm{W}}^s_0 \oplus 
 \widetilde{\bm{W}}^f_0 \oplus 
 {\bm{W}} \,.
\end{equation}
Using this decomposition in the monolithic weak form of the FSI problem,
we may define three operators.
We look at the domain as the subdivision of three overlapping regions:
the fluid domain, the solid domain 
and a domain around the fluid-solid interface.
 By gathering all the integrals in the respective domains,
we define the \textit{fluid operator} $ \mathcal{F} $, 
the  \textit{solid operator} $ \mathcal{S} $ and the
\textit{interface operator}  $ \mathcal{I} $ as
\begin{align}
 \mathcal{F} & : \bm{V} \times \bm{V} \times L^2(\Omega_t)
 \rightarrow 
 ({\bm{W}}^f_0 \times L^2(\Omega^{f}_t) \times {\bm{W}}^f_{0,\bm{d}^f} )'\,, \\
  \mathcal{S} & : \bm{V} \times \bm{V} \times L^2(\Omega_t)
 \rightarrow 
 ({\bm{W}}^s_0 \times L^2(\Omega^{s}_t) \times \bm{H}^1(\Omega^s_t)  )' \,, \\
 \mathcal{I} & : \bm{V} \times \bm{V} \times L^2(\Omega_t)
 \rightarrow 
 ( \bm{H}^{\frac{1}{2}}(\Gamma^i_t) ) ' \,.
 \end{align}
 In particular, the interface operator gathers all the integrals on a support around the interface
 that arise from the momentum equations. 
 Thus, we have 
 \begin{multline}
 < \mathcal{F} (\bm{d}, \bm{u}, p), (\bm{\phi}^f, \phi^{pf}, {\bm{\phi}}^{\bm{kf}}) > 
 \, := \,
 \\    
  \frac{d}{dt} \left( \rho^f \bm{u} , \bm{\phi}^f \right)_{\Omega^f_t }
 -  \left( \rho^f ( \nabla \cdot \frac{\partial \bm{d}}{\partial t} ) \bm{u}             , \bm{\phi}^f \right)_{\Omega^f_t}
 +  \left( \rho^f [ (\bm{u} - \frac{\partial \bm{d}}{\partial t} ) \cdot \nabla ] \bm{u} , \bm{\phi}^f \right)_{\Omega^f_t}  
\\ 
+  \left( \sigma^f (\bm{u},p)                                                    , \nabla \bm{\phi}^f \right)_{\Omega^f_t}
-  \left( \rho^f \bm{f}^f                                                               , \bm{\phi}^f \right)_{\Omega^f_t}
\\ 
+ \left( \nabla \cdot  \bm{u} , \phi^{pf} \right)_{{\Omega}^f_t}  
+ \left( k(\aled{\bm{{x}}}) ( \aled{\nabla} \aled{\bm{d}} + (\aled{\nabla} \aled{\bm{d}})^T ), \aled{\nabla} \widehat{\bm{\phi}}^{\bm{kf}}  \right)_{\widehat{\Omega}^f}
\end{multline}
for all $ (\bm{\phi}^f, \phi^{pf}, {\bm{\phi}}^{\bm{kf}}) \in 
{\bm{W}}^f_0 \times L^2(\Omega^{f}_t) \times {\bm{W}}^f_{0,\bm{d}^f} $,
\begin{multline}
 <  \mathcal{S} (\bm{d}, \bm{u}, p), (\bm{\phi}^s, \phi^{ps}, {\bm{\phi}}^{\bm{ks}}) > 
 \, := \,
 \\ 
  \left( \widehat{\rho}^s \widehat{J}( \widehat{\bm{d}} ) \frac{\partial \widehat{\bm{u}}}{\partial t} , \widehat{\bm{\phi}}^s         \right)_{\widehat{\Omega}^s}
 + \left( \widehat{J}( \widehat{\bm{d}} ) \widehat{\bm{\sigma}}^s(\widehat{\bm{d}},\widehat{p}) (\widehat{\bm{F}}(\widehat{\bm{d}}))^{-T}            , \widehat{\nabla} \widehat{\bm{\phi}}^s  \right)_{\widehat{\Omega}^s} 
 - \left( \widehat{\rho}^s \widehat{J}( \widehat{\bm{d}} ) \widehat{\bm{f}}^s    ,  \widehat{\bm{\phi}}^s        \right)_{\widehat{\Omega}^s} 
\\ 
 + \left( \widehat{J}( \widehat{\bm{d}} ) - 1 , \widehat{\phi}^{ps}  \right)_{\widehat{\Omega}^s} 
 + \left( \widehat{\bm{u}} - \frac{\partial \widehat{\bm{d}}}{\partial t} , \widehat{\bm{\phi}}^{\bm{ks}} \right)_{\widehat{\Omega}^s}
\end{multline}
for all $ (\bm{\phi}^s, \phi^{ps}, {\bm{\phi}}^{\bm{ks}})  \in 
{\bm{W}}^s_0 \times L^2(\Omega^{s}_t) \times \bm{H}^1(\Omega^s_t) $ and
\begin{equation}
< \mathcal{I} ( \bm{d}, \bm{u}, p), \bm{\mu} > 
 \, := \,
    < \mathcal{F} (\bm{d}, \bm{u}, p), ( \mathcal{L}^f_t \bm{\mu}, 0, \bm{0}) >
 +  < \mathcal{S} (\bm{d}, \bm{u}, p), ( \mathcal{L}^s_t \bm{\mu}, 0, \bm{0}) >  
\end{equation}
for all $ \bm{\mu} \in \bm{H}^{1/2}(\Gamma^i_t) $.
Therefore, the monolithic weak FSI problem is equivalent to the operator system
\begin{equation} \label{coupled_fsi_continuous_with_operators}
 \mathcal{F} (\bm{d}, \bm{u}, p) = 0 \,,\quad
 \mathcal{S} (\bm{d}, \bm{u}, p) = 0 \,, \quad
 \mathcal{I} (\bm{d}, \bm{u}, p) = 0 \,. 
\end{equation}

\subsection{Discretization of the operator formulation}

The weak FSI problem is now discretized in time and space.
 Different time derivatives are involved in the FSI system:
 the time derivative of the solid velocity in the solid momentum equation, 
 the time derivative of the fluid velocity 
 as well as that of the fluid domain displacement in the fluid momentum equation.
We describe the time-discretized formulation 
by using time-discretized counterparts $ \mathcal{F}_n $, $ \mathcal{S}_n $ and $ \mathcal{I}_n $ 
of the $ \mathcal{F} $, $ \mathcal{S} $ and $ \mathcal{I} $ operators.
We construct the scheme by using backward finite differences for the time derivatives in the integrands,
while we use a Crank-Nicolson scheme for the fluid and solid momentum balances.
The choice of the time discretization scheme is not a major focus of this work.
For more details, we refer to  \cite{TurekBenchmark2,wick2011fluid}.

Let $ 0 = t_0 < t_1 < ... < t_N = T $
be a subdivision of the time interval
 with constant time step $ \Delta t = t_{n+1} - t_{n} $.
We denote with $ \Omega^f_{n} $, $ \Omega^s_{n} $ and $ \Gamma^i_{n} $ the fluid domain, solid domain
and fluid-solid interface at time $ t_n $.
Let $(\bm{d}_n, \bm{u}_n, p_n)$ be the approximation at time $t_n$. 
For $ n = 0,1,...,N $,
the \textit{time-discretized fluid operator} $ \mathcal{F}_n $ is defined 
for all test functions $ (\bm{\phi}^f, \phi^{pf}, {\bm{\phi}}^{\bm{kf}} ) $ as

\begin{multline}
  < \mathcal{F}_n (\bm{d}_{n+1}, \bm{u}_{n+1}, p_{n+1}), (\bm{\phi}^f, \phi^{pf}, {\bm{\phi}}^{\bm{kf}} ) > 
 \, := \,
 \\    
  \frac{1}{\Delta t} \left( \rho^f \bm{u}_{n+1} , \bm{\phi}^f_{n+1} \right)_{\Omega^f_{n+1} }
- \frac{1}{\Delta t} \left( \rho^f \bm{u}_n, \bm{\phi}^f_{n} \right)_{\Omega^f_{n} } 
\\  
- \frac{1}{2}        \left( \rho^f ( \nabla \cdot \frac{(\bm{d}_{n+1} - \bm{d}_{n})}{\Delta t} ) \bm{u}_{n+1}           , \bm{\phi}^f_{n+1} \right)_{\Omega^f_{n+1}}
- \frac{1}{2}        \left( \rho^f ( \nabla \cdot \frac{(\bm{d}_{n+1} - \bm{d}_{n})}{\Delta t} ) \bm{u}_{n}             , \bm{\phi}^f_{n} \right)_{\Omega^f_{n}}
\\
+ \frac{1}{2}  \left( \rho^f [ (\bm{u}_{n+1} - \frac{(\bm{d}_{n+1} - \bm{d}_{n})}{\Delta t} ) \cdot \nabla ] \bm{u}_{n+1} , \bm{\phi}^f_{n+1} \right)_{\Omega^f_{n+1}}  
\\
+ \frac{1}{2}  \left( \rho^f [ (\bm{u}_{n}   - \frac{(\bm{d}_{n+1} - \bm{d}_{n})}{\Delta t} ) \cdot \nabla ] \bm{u}_{n}   , \bm{\phi}^f_{n}   \right)_{\Omega^f_{n}}  
\\
+ \frac{1}{2}  \left( \sigma^f (\bm{u}_{n+1},p_{n+1})                      , \nabla \bm{\phi}^f_{n+1} \right)_{\Omega^f_{n+1}}
+ \frac{1}{2}  \left( \sigma^f (\bm{u}_{n},p_{n})                          , \nabla \bm{\phi}^f_{n}   \right)_{\Omega^f_{n}}
\\
- \frac{1}{2}  \left( \rho^f \bm{f}^f_{n+1}                                , \bm{\phi}^f_{n+1} \right)_{\Omega^f_{n+1}}
- \frac{1}{2}  \left( \rho^f \bm{f}^f_{n}                                  , \bm{\phi}^f_{n}   \right)_{\Omega^f_{n}}
\\ 
+ \left( \nabla \cdot  \bm{u}_{n+1}, \phi^{pf} \right)_{{\Omega}^f_{n+1}}  
+ \left( k(\aled{\bm{{x}}}) ( \aled{\nabla} \aled{\bm{d}}_{n+1} + (\aled{\nabla} \aled{\bm{d}}_{n+1})^T ),   \aled{\nabla} \widehat{\bm{\phi}}^{\bm{kf}}  \right)_{\widehat{\Omega}^f}
 \,.
 \end{multline}

The \textit{time-discretized solid operator} $ \mathcal{S}_n $ is defined 
for all $ (\bm{\phi}^s, \phi^{ps}, {\bm{\phi}}^{\bm{ks}}) $ as 
\begin{multline}
 <  \mathcal{S}_{n} (\bm{d}_{n+1}, \bm{u}_{n+1}, p_{n+1}), (\bm{\phi}^s, \phi^{ps}, {\bm{\phi}}^{\bm{ks}}) > 
 \, := \,
 \\ 
   \frac{1}{\Delta t}  \left( \widehat{\rho}^s \widehat{J}( \widehat{\bm{d}}_{n+1} ) \widehat{\bm{u}}_{n+1} , \widehat{\bm{\phi}}^s         \right)_{\widehat{\Omega}^s}
 - \frac{1}{\Delta t}  \left( \widehat{\rho}^s \widehat{J}( \widehat{\bm{d}}_{n} )   \widehat{\bm{u}}_{n}   , \widehat{\bm{\phi}}^s         \right)_{\widehat{\Omega}^s}
 \\
 + \frac{1}{2} \left( \widehat{J}( \widehat{\bm{d}}_{n+1} ) \widehat{\bm{\sigma}}^s(\widehat{\bm{d}}_{n+1},\widehat{p}_{n+1}) (\widehat{\bm{F}}(\widehat{\bm{d}}_{n+1}))^{-T}     , \widehat{\nabla} \widehat{\bm{\phi}}^s  \right)_{\widehat{\Omega}^s} 
 \\
 + \frac{1}{2} \left( \widehat{J}( \widehat{\bm{d}}_{n} )   \widehat{\bm{\sigma}}^s(\widehat{\bm{d}}_{n},\widehat{p}_{n})     (\widehat{\bm{F}}(\widehat{\bm{d}}_{n}))^{-T}       , \widehat{\nabla} \widehat{\bm{\phi}}^s  \right)_{\widehat{\Omega}^s} 
\\
 - \frac{1}{2} \left( \widehat{\rho}^s \widehat{J}( \widehat{\bm{d}}_{n+1} ) \widehat{\bm{f}}^s_{n+1}    ,  \widehat{\bm{\phi}}^s        \right)_{\widehat{\Omega}^s} 
 - \frac{1}{2} \left( \widehat{\rho}^s \widehat{J}( \widehat{\bm{d}}_{n} )   \widehat{\bm{f}}^s_{n}      ,  \widehat{\bm{\phi}}^s        \right)_{\widehat{\Omega}^s} 
\\ 
 + \left( \widehat{J}( \widehat{\bm{d}}_{n+1} ) - 1 , \widehat{\phi}^{ps}  \right)_{\widehat{\Omega}^s} 
\\ 
 - \frac{1}{\Delta t} \left( \widehat{\bm{d}}_{n+1} , \widehat{\bm{\phi}}^{\bm{ks}} \right)_{\widehat{\Omega}^s}
 + \frac{1}{\Delta t} \left( \widehat{\bm{d}}_{n}   , \widehat{\bm{\phi}}^{\bm{ks}} \right)_{\widehat{\Omega}^s}
\\
 + \frac{1}{2} \left( \widehat{\bm{u}}_{n+1} , \widehat{\bm{\phi}}^{\bm{ks}} \right)_{\widehat{\Omega}^s}
 + \frac{1}{2} \left( \widehat{\bm{u}}_{n}   , \widehat{\bm{\phi}}^{\bm{ks}} \right)_{\widehat{\Omega}^s}
 \,.
\end{multline}

Finally, the \textit{time-discretized interface operator}  $ \mathcal{I}_n $ is defined
for all $ \bm{\mu} $ as 
\begin{multline*}
< \mathcal{I}_{n} ( \bm{d}_{n+1}, \bm{u}_{n+1}, p_{n+1}), \bm{\mu} > 
:=
\\
 \, = \,
    < \mathcal{F}_{n} (\bm{d}_{n+1}, \bm{u}_{n+1}, p_{n+1}), ( \mathcal{L}^f_t \bm{\mu}, 0, \bm{0}) >
 +  < \mathcal{S}_{n} (\bm{d}_{n+1}, \bm{u}_{n+1}, p_{n+1}), ( \mathcal{L}^s_t \bm{\mu}, 0, \bm{0}) >  \,.
\end{multline*}
Therefore, the time-discretized weak FSI problem 
\eqref{coupled_fsi_continuous_with_operators} consists in
finding the iterates $(\bm{d}_{n+1}, \bm{u}_{n+1}, p_{n+1}) $ for $ n = 0, 1, ..., N $ such that 
\begin{equation}  \label{coupled_fsi_semi-discretized_weak_operators}
 \begin{cases}
 \mathcal{F}_n (\bm{d}_{n+1}, \bm{u}_{n+1}, p_{n+1}) = 0 \,,\\
 \mathcal{S}_n (\bm{d}_{n+1}, \bm{u}_{n+1}, p_{n+1}) = 0 \,, \\
 \mathcal{I}_n (\bm{d}_{n+1}, \bm{u}_{n+1}, p_{n+1}) = 0 \,. \\
 \end{cases}
\end{equation}


For every open set $ \mathcal{O} \subset \mathbb{R}^n $, let $ \mathcal{O}_h $ be a 
\review{regular and geometrically conforming} triangulation of $ \mathcal{O} $
made of quadrilaterals when $ n = 2 $ or hexahedra when $ n = 3 $.
For the sake of simplicity, we assume that $ \mathcal{O} $ has polygonal (or polyhedral) boundary.
Furthermore, we assume that the fluid-solid interface 
coincides with a subset of adjacent edges in the triangulation.
We denote by $ {\Phi}(\mathcal{O}_h) \subset {H}^1(\mathcal{O})$
the space of scalar-valued continuous piecewise-biquadratic (or triquadratic) polynomials
and by 
$ \Psi(\mathcal{O}_h) \subset L^2(\mathcal{O}) $ the space of discontinuous piecewise-linear polynomials, i.e.
 \begin{align*}
 {\Phi}(\mathcal{O}_h) & = \{ {\phi} \in \mathcal{C}^0(\overline{\mathcal{O}}_h) \, : 
\, {\phi}|_{\kappa} \in Q_2(\kappa) \quad \forall \,   \kappa \in \mathcal{O}_h \} \,, \\
 \Psi(\mathcal{O}_h)  & = \{ \psi \in L^2(\mathcal{O}_h) \, : 
\, \psi|_{\kappa} \in P_1(\kappa) \quad \forall \,   \kappa \in \mathcal{O}_h \} \,.
\end{align*}
In order to deal with Dirichlet conditions, 
we consider the subspaces with zero trace on the Dirichlet parts of the boundary,
and we make a little abuse of notation by keeping the same symbols for these subspaces.
Also, we denote $ n $ copies of $ {\Phi}(\mathcal{O}_h) $ with the boldface notation $ \bm{\Phi}(\mathcal{O}_h) $.
The pair of finite element spaces $ ( \bm{\Phi}(\mathcal{O}_h) , \Psi(\mathcal{O}_h) ) $ 
\review{is referred to as the Q2P1(disc) pair 
and it belongs to a Crouzeix-Raviart family \cite{cuvelier1986finite,crouzeix1973conforming,GiraultRaviart}}. 
We remark that this pair satisfies the discrete inf-sup condition, 
so that the divergence-free constraint can be enforced without stabilization terms,
 both for the incompressible elasticity equations 
 and for the incompressible Navier-Stokes equations.
For other works on FSI using the \review{Q2P1(disc) pair}, see \cite{TurekBenchmark2,wick2011fluid}. 
We also remark that we use the \review{Q2P1(disc) pair} both in two and three dimensions, unlike \cite{TurekBenchmark2}, 
where a different inf-sup stable pair is chosen in the 3D case.
 \review{We point out that the use of mixed finite elements
 for the incompressible elasticity equations is one of the methods to avoid locking effects \cite{wihler2006locking}.
 In this respect, we observed a good behaviour of our solver in preliminary numerical results.
 }

The space of traces over the interface $ \widehat{\Gamma}^i_h $ 
of functions in $ \bm{\Phi}(\widehat{\Omega}_h) $
is denoted as $ \Pi_h(\widehat{\Gamma}^i_h) $.
Given the time-discretized operators, we denote their fully discretized versions 
as $  \mathcal{F}_{n,h} $,  $  \mathcal{S}_{n,h} $ and  $  \mathcal{I}_{n,h} $.
They are defined as
\begin{align*}
&  \mathcal{F}_{n,h} :  \bm{\Phi}(\Omega_{t,h}) \times \bm{\Phi}(\Omega_{t,h}) \times \Psi(\Omega_{t,h})
  \rightarrow \\
& \rightarrow
  ( \bm{\Phi}(\Omega^{f}_{t,h}) \cap \bm{H}^1_0(\Omega^{f}_{t,h}  ) \times 
 \Psi(\Omega^{f}_{t,h}) \times 
 ( \bm{\Phi} (\Omega^f_{t,h}) \cap  \bm{H}^1_0(\Omega^{f}_{t,h}) )
 )' 
 \,, 
 \\
&  \mathcal{S}_{n,h} :  \bm{\Phi}(\Omega_{t,h}) \times \bm{\Phi}(\Omega_{t,h}) \times \Psi(\Omega_{t,h})
  \rightarrow \\
& \rightarrow
  ( \bm{\Phi}(\Omega^{s}_{t,h}) \cap  \bm{H}^1_0(\Omega^{s}_{t,h})  \times \Psi(\Omega^{s}_{t,h}) \times \bm{\Phi} (\Omega^s_{t,h})  )' 
 \,, 
 \\
& \mathcal{I}_{n,h} : \bm{\Phi}(\Omega_{t,h}) \times \bm{\Phi}(\Omega_{t,h}) \times \Psi(\Omega_{t,h})
 \rightarrow 
 ( \Pi_h(\widehat{\Gamma}^i_h) ) '
 \,, 
\end{align*}
we have
\begin{equation}
\begin{aligned}
 \mathcal{F}_{n,h}(\bm{d}_h,\bm{u}_h,p_h ) & :=  \mathcal{F}_{n}(\bm{d}_h,\bm{u}_h,p_h ) \,,
 \\ 
 \mathcal{S}_{n,h}(\bm{d}_h,\bm{u}_h,p_h ) & :=  \mathcal{S}_{n}(\bm{d}_h,\bm{u}_h,p_h ) \,,
 \\ 
 \mathcal{I}_{n,h}(\bm{d}_h,\bm{u}_h,p_h ) & :=  \mathcal{I}_{n}(\bm{d}_h,\bm{u}_h,p_h ) \,.
 \end{aligned}
\end{equation}

The fully discrete version of the FSI problem 
\eqref{coupled_fsi_semi-discretized_weak_operators} 
 consists in searching for the variables
 $ (\bm{d}_{n+1,h}, \bm{u}_{n+1,h}, p_{n+1,h}) $ such that

\begin{equation}
 \label{coupled_fsi_semi-discretized_weak_operators_fem}
 \begin{cases}
 \mathcal{F}_{n,h} (\bm{d}_{n+1,h}, \bm{u}_{n+1,h}, p_{n+1,h}) = 0 \,,\\
 \mathcal{S}_{n,h} (\bm{d}_{n+1,h}, \bm{u}_{n+1,h}, p_{n+1,h}) = 0 \,, \\
 \mathcal{I}_{n,h} (\bm{d}_{n+1,h}, \bm{u}_{n+1,h}, p_{n+1,h}) = 0 \,. 
 \end{cases}
\end{equation}

With the given choice of the finite element space for the discretized displacement and velocity $ \bm{d}_h $ and $ \bm{u}_h $,
their continuity on the interface
is enforced strongly as this is a $ \mathcal{C}^0$ finite element space.
On the other hand, pressure continuity is not required across the solid-fluid interface.

%
%

\section{Monolithic Newton-Krylov solver with Multigrid-Richardson-Schwarz preconditioner}

This is the main section which describes the solver.
 The outer loop consists of an exact Newton linearization implemented by automatic differentiation \cite{hoganadept}.
  The solution of the linear systems is performed using a GMRES solver \cite{Saad} 
  preconditioned by a geometric multigrid V-cycle algorithm.
 The smoother is of modified Richardson type, 
 in turn preconditioned by a restricted additive Schwarz method.
 The coarse grid correction problem is dealt with
  by a direct solver of the monolithic system. 
  In the steady-state case, the previous procedure is wrapped by 
  a Full Multigrid algorithm \cite{Brenner}.
  More details on the various parameters 
  will be provided in the discussion on the numerical results.
  
  
\subsection{Structure of the Jacobian}

\begin{figure}[ht]
  \centering
  \includegraphics[width=0.4\textwidth]{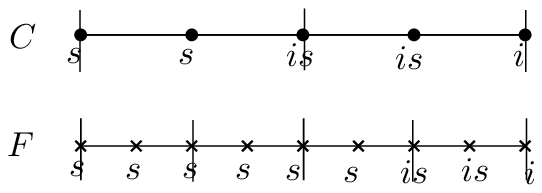} \hspace{1cm}
  \includegraphics[width=0.4\textwidth]{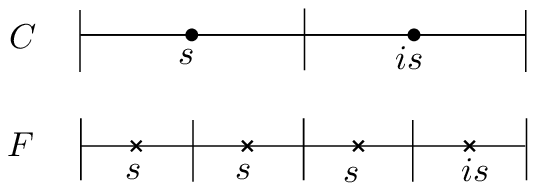}
  \caption{Node-based (left) and element-based (right) dof indices for the solid part, 
  and relationship between coarse (C) and fine (F) adjacent levels.
  A 1D representation is used here for the sake of simplicity.
  The same description holds for the fluid part.
  Notice that the region adjacent to the interface $i$  (characterized by the $is$ indices) restricts with refinement.
  Thus, certain degrees of freedom that belong to the ``interface region'' $is$ at a coarser level 
  belong to the ``bulk'' regions $s$ at the next finer level.}
    \label{domain_parts}
\end{figure}
 
Let $ \bm{y} = (\bm{d}, \bm{u}, p) $ denote the vector of unknown degrees of freedom and 
for every nonlinear step $ k $  denote the error  
as $ \bm{e}^{(k)} = \bm{y}^{(k)} - \bm{y}^{(k-1)} $.
At each nonlinear step a linear system $ \bm{J}^{(k)} \bm{e}^{(k+1)} = - \bm{r}^{(k)} $ is solved for $ \bm{e}^{(k+1)} $.
In order to describe the block structure of the exact Jacobian $ \bm{J}^{(k)} $ 
we divide the degrees of freedom  of the monolithic fields using the indices $ s $,  $ f $, $ i $, $ is $, $ if $, 
see Fig. \ref{domain_parts}.
The indices $ s $ and $ f $ are for the solid and fluid parts.
The index $ i $ is for the degrees of freedom associated to the interface.
The indices $ is $ and $ if $ indicate the degrees of freedom on the support adjacent to the interface but not on the interface,
and also belonging to the support around the interface on the solid side and fluid side, respectively.
Notice that we do not have interface pressure degrees of freedom 
because we use element-based degrees of freedom for it.
In symbols, the unknown vector is
\begin{equation*}
\begin{bmatrix}
\bm{d}^{s}     \,
\bm{d}^{is}    \, 
\bm{d}^{i}     \,
\bm{d}^{if}    \,
\bm{d}^{f}     \,
\vline         \,
\bm{u}^{s}     \,
\bm{u}^{is}    \,
\bm{u}^{i}     \,
\bm{u}^{if}    \,
\bm{u}^{f}     \,
\vline         \,
p^{s}          \,
p^{is}         \,
p^{if}         \,
p^{f} 
\end{bmatrix}^T\,.
\end{equation*}

Corresponding to this layout, the Jacobian has the block structure
\begin{equation}
\bm{J}^{(k)} =
\scalemath{0.75}{
\begin{bmatrix}[ ccccc || ccccc || cccc]
  S_{\bm{d}^s}^{\bm{d}^s}  &   S_{\bm{d}^{is}}^{\bm{d}^s} & 0             &   0              &       0        & S_{\bm{u}^s}^{\bm{d}^s}     & S_{\bm{u}^{is}}^{\bm{d}^s}    &      0           &        0 &     0    &   S_{p^s}^{\bm{d}^s}     &      0      &   0         &     0         \\
   S_{\bm{d}^s}^{\bm{d}^{is}}  &   S_{\bm{d}^{is}}^{\bm{d}^{is}} & S_{\bm{d}^{i}}^{\bm{d}^{is}}   &   0              &       0        & S_{\bm{u}^s}^{\bm{d}^{is}}     & S_{\bm{u}^{is}}^{\bm{d}^{is}}    & S_{\bm{u}^{i}}^{\bm{d}^{is}}   &        0 &     0    &   S_{p^{s}}^{\bm{d}^{is}}    &  S_{p^{is}}^{\bm{d}^{is}} &   0         &     0         \\
              0  &   I_{\bm{d}^{is}}^{\bm{d}^{i}} & I_{\bm{d}^{i}}^{\bm{d}^{i}}   & I_{\bm{d}^{if}}^{\bm{d}^{i}}  &       0        &         0        &   I_{\bm{u}^{is}}^{\bm{d}^{i}}  & I_{\bm{u}^{i}}^{\bm{d}^{i}}   & I_{\bm{u}^{if}}^{\bm{d}^{i}}  &     0    &   0          &  I_{p^{is}}^{\bm{d}^{i}} &   I_{p^{if}}^{\bm{d}^{i}}   &     0         \\
   0             &       0           & F_{\bm{d}^{i}}^{\bm{u}^{if}}   & F_{\bm{d}^{if}}^{\bm{u}^{if}}  & F_{\bm{d}^f}^{\bm{u}^{if}}   &         0        &         0          & F_{\bm{u}^{i}}^{\bm{u}^{if}}   & F_{\bm{u}^{if}}^{\bm{u}^{if}}  &  F_{\bm{u}^f }^{\bm{u}^{if}}  &   0          &     0    &   F_{p^{if}}^{\bm{u}^{if}}   &   F_{p^f}^{\bm{u}^{if}}      \\
   0             &       0           & 0                              & F_{\bm{d}^{if}}^{\bm{u}^f }    & F_{\bm{d}^f}^{\bm{u}^f }   &         0        &         0          &       0   & F_{\bm{u}^{if}}^{\bm{u}^f }  &  F_{\bm{u}^f }^{\bm{u}^f }  &   0          &     0    &   0   &   F_{p^f}^{\bm{u}^f }      \\
\hline
\hline
   K_{\bm{d}^s}^{\bm{u}^s}  &   K_{\bm{d}^{is}}^{\bm{u}^s} &           0   &   0              &   0            &    K_{\bm{u}^s}^{\bm{u}^s}  &   K_{\bm{u}^{is}}^{\bm{u}^s}  &   0          &    0     &     0    &   0          &     0    &      0      &     0         \\
   K_{\bm{d}^s}^{\bm{u}^{is}}  &   K_{\bm{d}^{is}}^{\bm{u}^{is}} & K_{\bm{d}^{i}}^{\bm{u}^{is}}   &   0              &   0            &    K_{\bm{u}^s}^{\bm{u}^{is}}  &   K_{\bm{u}^{is}}^{\bm{u}^{is}}  & K_{\bm{u}^{i}}^{\bm{u}^{is}}   &    0     &     0    &   0          &     0    &      0      &     0         \\
   0  &   K_{\bm{d}^{is}}^{\bm{u}^{i}} & K_{\bm{d}^{i}}^{\bm{u}^{i}}   &   {0}              &   0            &    0  &   K_{\bm{u}^{is}}^{\bm{u}^{i}}  & K_{\bm{u}^{i}}^{\bm{u}^{i}}   &   {0}     &     0    &   0          &     0    &      0      &     0         \\
              0  &       0           & A_{\bm{d}^{i}}^{\bm{d}^{if}}   & A_{\bm{d}^{if}}^{\bm{d}^{if}}  &  A_{\bm{d}^f}^{\bm{d}^{if}}  &         0        &         0          &     0    &    0     &     0    &   0          &     0    &      0      &     0         \\
              0  &       0           & 0   & A_{\bm{d}^{if}}^{\bm{d}^f}  &  A_{\bm{d}^f}^{\bm{d}^f}  &         0        &         0          &     0    &    0     &     0    &   0          &     0    &      0      &     0         \\
\hline
\hline
   V_{\bm{d}^s}^{p^s} &  V_{\bm{d}^{is}}^{p^s} &           0  &   0              &   0            &    0             &         0          &       0  &    0     &     0    &   0          &     0    &      0      &     0         \\
   0                  &  V_{\bm{d}^{is}}^{p^{is}} & V_{\bm{d}^{i}}^{p^{is}}  &   0              &   0            &    0             &         0          &       0  &    0     &     0    &   0          &     0    &      0      &     0         \\
   0             &       0           & W_{\bm{d}^{i}}^{p^{if}}  & W_{\bm{d}^{if}}^{p^{if}} &       0       &         0        &         0          & W_{\bm{u}^{i}}^{p^{if}}  & W_{\bm{u}^{if}}^{p^{if}} &    0     &   0          &     0    &      0      &     0      \\
   0             &       0           & 0   & W_{\bm{d}^{if}}^{p^{f}} & W_{\bm{d}^f}^{p^{f}}  &         0       &         0          &   0  & W_{\bm{u}^{if}}^{p^{f}} &  W_{\bm{u}^f }^{p^{f}} &   0          &     0    &      0      &     0      \\
\end{bmatrix}
}
\,.
\end{equation}

For each block, the superscript and subscript indicate the quantity 
associated to the row and column degrees of freedom, respectively.
The blocks associated to the operator $ \mathcal{F} $ are $F$, $ A $ and $W$,
which correspond to the fluid momentum, kinematic and continuity equations, respectively.
The blocks associated to the operator $ \mathcal{S} $ are $ S $, $K$ and $V$,
related to the solid momentum, kinematic and continuity balances.
The operator $ \mathcal{I} $ for the interface momentum has associated blocks $ I $.
 
Notice that while the block row associated to the interface displacements $ \bm{d}^i $,
which corresponds to the interface momentum balance,
 receives contributions both from the solid and from the fluid part,
 the block row related to the interface velocities $ \bm{u}^i $
 only receives contributions from the solid part.
 This corresponds to having zero blocks 
 immediately to the right of $ K_{\bm{d}^{i}}^{\bm{u}^{i}} $ and $ K_{\bm{u}^{i}}^{\bm{u}^{i}} $.
 In fact, the velocity at the interface is determined 
 by the solid kinematic equation only.
 
Some blocks would normally be zero in the case of fixed fluid domain, but they exist because 
they act on the fluid domain displacement, which determines the position 
of the mesh points in the fluid part. 
In the interface momentum equation, these are
$ I_{\bm{d}^{i}}^{\bm{d}^{i}} $    $ I_{\bm{d}^{if}}^{\bm{d}^{i}} $ 
 (the block $ I_{\bm{d}^{i}}^{\bm{d}^{i}} $ would actually be nonzero even with a fixed domain,
 as it receives contributions from the solid momentum);
in the fluid momentum we have 
$ F_{\bm{d}^{i}}^{\bm{u}^{if}} $, $ F_{\bm{d}^{if}}^{\bm{u}^{if}} $, $ F_{\bm{d}^f}^{\bm{u}^{if}} $  
 $ F_{\bm{d}^{if}}^{\bm{u}^f } $ and $ F_{\bm{d}^f}^{\bm{u}^f } $,
while in the fluid continuity we have 
$ W_{\bm{d}^{i}}^{p^{if}} $, $ W_{\bm{d}^{if}}^{p^{if}} $, 
$ W_{\bm{d}^{if}}^{p^{f}} $ and $ W_{\bm{d}^f}^{p^{f}} $. 

  We remark that in the steady-state case certain blocks 
  differ from the time-dependent case. 
  Among these, the blocks that become zero are:
  $ S_{\bm{u}^s}^{\bm{d}^s}     $,
  $ S_{\bm{u}^{is}}^{\bm{d}^s}  $,
  $ S_{\bm{u}^s}^{\bm{d}^{is}} $, $ S_{\bm{u}^{is}}^{\bm{d}^{is}}  $ and $ S_{\bm{u}^{i}}^{\bm{d}^{is}} $,    
        for zero time derivative in the solid momentum;
    $ K_{\bm{d}^s}^{\bm{u}^s} $, $  K_{\bm{d}^{is}}^{\bm{u}^s} $, 
    $ K_{\bm{d}^s}^{\bm{u}^{is}} $, $  K_{\bm{d}^{is}}^{\bm{u}^{is}} $, $ K_{\bm{d}^{i}}^{\bm{u}^{is}} $ 
    $ K_{\bm{d}^{is}}^{\bm{u}^{i}} $ and $ K_{\bm{d}^{i}}^{\bm{u}^{i}} $,  
        for zero time derivative in the solid kinematic equation;
 $ I_{\bm{u}^{is}}^{\bm{d}^{i}} $, for the zero time derivative in the interface momentum
(the block $ I_{\bm{u}^{i}}^{\bm{d}^{i}} $ is filled from the fluid part anyway).

We point out that a better computational performance was observed by switching the $F$ block rows with the $A$ block rows, namely
\begin{equation} \label{jac_blocks_shuf}
\bm{J}^{(k)}_{shuf} =
\scalemath{0.75}{
\begin{bmatrix}[ ccccc || ccccc || cccc]
  S_{\bm{d}^s}^{\bm{d}^s}  &   S_{\bm{d}^{is}}^{\bm{d}^s} & 0             &   0              &       0        & S_{\bm{u}^s}^{\bm{d}^s}     & S_{\bm{u}^{is}}^{\bm{d}^s}    &      0           &        0 &     0    &   S_{p^s}^{\bm{d}^s}     &      0      &   0         &     0         \\
   S_{\bm{d}^s}^{\bm{d}^{is}}  &   S_{\bm{d}^{is}}^{\bm{d}^{is}} & S_{\bm{d}^{i}}^{\bm{d}^{is}}   &   0              &       0        & S_{\bm{u}^s}^{\bm{d}^{is}}     & S_{\bm{u}^{is}}^{\bm{d}^{is}}    & S_{\bm{u}^{i}}^{\bm{d}^{is}}   &        0 &     0    &   S_{p^{s}}^{\bm{d}^{is}}    &  S_{p^{is}}^{\bm{d}^{is}} &   0         &     0         \\
              0  &   I_{\bm{d}^{is}}^{\bm{d}^{i}} & I_{\bm{d}^{i}}^{\bm{d}^{i}}   & I_{\bm{d}^{if}}^{\bm{d}^{i}}  &       0        &         0        &   I_{\bm{u}^{is}}^{\bm{d}^{i}}  & I_{\bm{u}^{i}}^{\bm{d}^{i}}   & I_{\bm{u}^{if}}^{\bm{d}^{i}}  &     0    &   0          &  I_{p^{is}}^{\bm{d}^{i}} &   I_{p^{if}}^{\bm{d}^{i}}   &     0         \\
              0  &       0           & A_{\bm{d}^{i}}^{\bm{d}^{if}}   & A_{\bm{d}^{if}}^{\bm{d}^{if}}  &  A_{\bm{d}^f}^{\bm{d}^{if}}  &         0        &         0          &     0    &    0     &     0    &   0          &     0    &      0      &     0         \\
              0  &       0           & 0   & A_{\bm{d}^{if}}^{\bm{d}^f}  &  A_{\bm{d}^f}^{\bm{d}^f}  &         0        &         0          &     0    &    0     &     0    &   0          &     0    &      0      &     0         \\
\hline
\hline
   K_{\bm{d}^s}^{\bm{u}^s}  &   K_{\bm{d}^{is}}^{\bm{u}^s} &           0   &   0              &   0            &    K_{\bm{u}^s}^{\bm{u}^s}  &   K_{\bm{u}^{is}}^{\bm{u}^s}  &   0          &    0     &     0    &   0          &     0    &      0      &     0         \\
   K_{\bm{d}^s}^{\bm{u}^{is}}  &   K_{\bm{d}^{is}}^{\bm{u}^{is}} & K_{\bm{d}^{i}}^{\bm{u}^{is}}   &   0              &   0            &    K_{\bm{u}^s}^{\bm{u}^{is}}  &   K_{\bm{u}^{is}}^{\bm{u}^{is}}  & K_{\bm{u}^{i}}^{\bm{u}^{is}}   &    0     &     0    &   0          &     0    &      0      &     0         \\
   0  &   K_{\bm{d}^{is}}^{\bm{u}^{i}} & K_{\bm{d}^{i}}^{\bm{u}^{i}}   &   {0}              &   0            &    0  &   K_{\bm{u}^{is}}^{\bm{u}^{i}}  & K_{\bm{u}^{i}}^{\bm{u}^{i}}   &   {0}     &     0    &   0          &     0    &      0      &     0         \\
   0             &       0           & F_{\bm{d}^{i}}^{\bm{u}^{if}}   & F_{\bm{d}^{if}}^{\bm{u}^{if}}  & F_{\bm{d}^f}^{\bm{u}^{if}}   &         0        &         0          & F_{\bm{u}^{i}}^{\bm{u}^{if}}   & F_{\bm{u}^{if}}^{\bm{u}^{if}}  &  F_{\bm{u}^f }^{\bm{u}^{if}}  &   0          &     0    &   F_{p^{if}}^{\bm{u}^{if}}   &   F_{p^f}^{\bm{u}^{if}}      \\
   0             &       0           & 0                              & F_{\bm{d}^{if}}^{\bm{u}^f }    & F_{\bm{d}^f}^{\bm{u}^f }   &         0        &         0          &       0   & F_{\bm{u}^{if}}^{\bm{u}^f }  &  F_{\bm{u}^f }^{\bm{u}^f }  &   0          &     0    &   0   &   F_{p^f}^{\bm{u}^f }      \\
\hline
\hline
   V_{\bm{d}^s}^{p^s} &  V_{\bm{d}^{is}}^{p^s} &           0  &   0              &   0            &    0             &         0          &       0  &    0     &     0    &   0          &     0    &      0      &     0         \\
   0                  &  V_{\bm{d}^{is}}^{p^{is}} & V_{\bm{d}^{i}}^{p^{is}}  &   0              &   0            &    0             &         0          &       0  &    0     &     0    &   0          &     0    &      0      &     0         \\
   0             &       0           & W_{\bm{d}^{i}}^{p^{if}}  & W_{\bm{d}^{if}}^{p^{if}} &       0       &         0        &         0          & W_{\bm{u}^{i}}^{p^{if}}  & W_{\bm{u}^{if}}^{p^{if}} &    0     &   0          &     0    &      0      &     0      \\
   0             &       0           & 0   & W_{\bm{d}^{if}}^{p^{f}} & W_{\bm{d}^f}^{p^{f}}  &         0       &         0          &   0  & W_{\bm{u}^{if}}^{p^{f}} &  W_{\bm{u}^f }^{p^{f}} &   0          &     0    &      0      &     0      \\
\end{bmatrix}
}
\,.
\end{equation}
Clearly, this affects the block structure of the subdomain matrices extracted from the Jacobian in the Schwarz algorithm, 
as will be described in the following.
As underlined in \cite{barker2010}, different orderings of equations and unknowns, though equivalent mathematically,
can have a significant effect on the convergence properties and computational time of the solver, 
especially in the parallel setting. The authors in \cite{barker2010} ordered the equations element by element.
In our work we followed a field-ordering approach as in \cite{fernandez2005newton}.

\subsection{Geometric Multigrid preconditioner}

As a preconditioner to the outer monolithic GMRES iteration,
we consider the action of geometric multigrid.
It is known that the condition number of the FSI Jacobian is very large,
as will be shown in Section \ref{sec_numres}.
Consider $ L $ levels of triangulations $ \Omega_{h_l} $ with associated mesh size $ h_l $
obtained recursively by simple midpoint refinement 
from an original geometrically conforming coarse triangulation $ \Omega_{h_0} $.
The finite element spaces associated to each level triangulation $ \Omega_{h_l} $
are $ \bm{\Phi}({\Omega}_{h_{l}}) $ and $ {\Psi}({\Omega}_{h_{l}}) $.
 The \textit{prolongation} $I^{l}_{l-1}$ and \textit{restriction} $I_{l}^{l-1}$ operators
 are defined as 
\begin{align}
 I_{l-1}^{l} & :
 \bm{\Phi}({\Omega}_{h_{l-1}}) \times 
 \bm{\Phi}({\Omega}_{h_{l-1}}) \times 
 {\Psi}({\Omega}_{h_{l-1}})
 \rightarrow 
 \bm{\Phi}({\Omega}_{h_{l}}) \times 
 \bm{\Phi}({\Omega}_{h_{l}}) \times 
 {\Psi}({\Omega}_{h_{l}})   \,, 
 \\
 I_{l}^{l-1} & :
 \bm{\Phi}({\Omega}_{h_{l}}) \times 
 \bm{\Phi}({\Omega}_{h_{l}}) \times 
 {\Psi}({\Omega}_{h_{l}})   
 \rightarrow 
 \bm{\Phi}({\Omega}_{h_{l-1}}) \times 
 \bm{\Phi}({\Omega}_{h_{l-1}}) \times 
 {\Psi}({\Omega}_{h_{l-1}}) \,,  
 \\
  I_{l-1}^l \bm{v} & = \bm{v} , \quad (I^{l-1}_l \bm{w} , \bm{v})  = (\bm{w} , I_{l-1}^l \bm{v}) \,,
\end{align}
for all $ \bm{ v } \in \bm{\Phi}({\Omega}_{h_{l-1}}) \times \bm{\Phi}({\Omega}_{h_{l-1}}) 
\times {\Psi}({\Omega}_{h_{l-1}}) $ and   
 $ \bm{ w } \in \bm{\Phi}({\Omega}_{h_{l}}) \times \bm{\Phi}({\Omega}_{h_{l}}) \times {\Psi}({\Omega}_{h_{l}}) $. 
 The prolongation is the natural injection from the coarse to the fine space,
 while the restriction operator $ {I}_{l}^{l-1} $ is the adjoint
of $ {I}^{l}_{l-1} $ with respect to the $ L^2 $ inner product \cite{braess2007finite,Brenner}.
Once finite element bases are chosen, 
 the matrix representations of the prolongation and restriction operators 
 will be denoted with the boldface notations $ \bm{I}_{l-1}^{l} $ and $ \bm{I}_{l}^{l-1} $.
Using these intergrid matrices, the coarse Jacobian $ \bm{J}_{l-1} $ can be computed from the fine Jacobian $ \bm{J}_l $ as
\begin{equation}
   \bm{J}_{l-1} = \bm{I}_{l}^{l-1} \bm{J}_{l} \bm{I}^{l}_{l-1} \,.
\end{equation}
The multigrid operators require the appropriate matrix representation
in order to enforce correctly the interface displacement and velocity continuity conditions
 at each level. They are given by 
\begin{equation} \label{restr_blocks}
 \bm{I}_{l}^{l-1} =
\scalemath{0.75}{
\begin{bmatrix}[ ccccc || ccccc || cccc]
   R_{\bm{d}^s}^{\bm{d}^s}          &    0                                  &      0                                        &   0              &       0        & 0     & 0    &      0           &        0 &     0    &   0     &      0      &   0         &     0         \\
    R_{\bm{d}^{s}}^{\bm{d}^{is}}     &    R_{\bm{d}^{is}}^{\bm{d}^{is}}      &    0     &   0              &       0        &  0     & 0    & 0  &        0 &     0    &   0    &  0   &   0         &     0         \\
     R_{\bm{d}^{s}}^{\bm{d}^{i}}    &    R_{\bm{d}^{is}}^{\bm{d}^{i}}    & R_{\bm{d}^{i}}^{\bm{d}^{i}}   &   R_{\bm{d}^{if}}^{\bm{d}^{i}}      &  R_{\bm{d}^{f}}^{\bm{d}^{i}}        &         0        &   0    &    0   &   0     &  0    &   0          &  0    &   0   &     0   \\
                                        0  &       0    &     0   & R_{\bm{d}^{if}}^{\bm{d}^{if}}  &   R^{\bm{d}^{if}}_{\bm{d}^f}      &         0        &         0          &     0    &    0     &     0    &   0          &     0    &      0      &     0         \\
              0  &       0           &  0   &   0    &  R_{\bm{d}^f}^{\bm{d}^f}  &         0        &         0          &     0    &    0     &     0    &   0          &     0    &      0      &     0         \\
\hline
\hline
   0  &   0   &           0   &   0              &   0            &    R_{\bm{u}^s}^{\bm{u}^s}  &   0     &     0   &    0     &     0    &   0          &     0    &      0      &     0         \\
   0  &   0 & 0   &   0              &   0            &    R^{\bm{u}^{is}}_{\bm{u}^s}  &   R_{\bm{u}^{is}}^{\bm{u}^{is}}  & 0  &    0     &     0    &   0          &     0    &      0      &     0         \\
   0  &   0 & 0   &   0              &   0            &    R^{\bm{u}^{i}}_{\bm{u}^s}  &   R^{\bm{u}^{i}}_{\bm{u}^{is}}   & R_{\bm{u}^{i}}^{\bm{u}^{i} }  &    0     &   0    &   0          &     0    &      0      &     0         \\
   0             &       0           &   0   & 0  &    0   &         0        &         0          &    0   & R_{\bm{u}^{if}}^{\bm{u}^{if}}  &  R^{\bm{u}^{if} }_{\bm{u}^f }  &   0          &     0    &   0   &   0      \\
   0             &       0           & 0   &  0   &   0    &         0        &         0          &       0    &    0  &  R_{\bm{u}^f }^{\bm{u}^f }  &   0          &     0    &   0   &   0      \\
\hline
\hline
   0             &  0   &           0  &   0              &   0            &    0             &         0          &       0  &    0     &     0              &     R_{p^s}^{p^s} & 0    &      0      &     0         \\
   0             &  0  &   0              &   0            &    0             &         0        &      0   &       0  &    0     &     0    &    R^{p^{is}}_{p^s}          &     R_{p^{is}}^{p^{is}} &      0      &     0         \\
   0             &       0           & 0   &       0       &         0       &      0 &      0  &         0          & 0  &    0     &   0          &     0    &      R_{p^{if}}^{p^{if}}      &     R^{p^{if}}_{p^{f}}      \\
   0             &       0           & 0   & 0 & 0   &         0       &         0          &   0  & 0   &  0 &   0          &     0    &     0      &       R_{p^{f}}^{p^{f}} \\
\end{bmatrix}
}
\,,
\end{equation}

\begin{equation} 
\setlength{\tabcolsep}{1pt}
\renewcommand{\arraystretch}{0.5}
\label{prol_blocks}
 \bm{I}_{l-1}^{l} =
\scalemath{0.75}{
\begin{bmatrix}[ ccccc || ccccc || cccc]
   P_{\bm{d}^s}^{\bm{d}^s}        &    P_{\bm{d}^{is}}^{\bm{d}^{s}}  &  0   &   0              &       0        & 0     & 0    &      0           &        0 &     0    &   0     &      0      &   0         &     0         \\
   0  &    P_{\bm{d}^{is}}^{\bm{d}^{is}} &    P_{\bm{d}^{i}}^{\bm{d}^{is}}  &   0              &       0        &  0     & 0    & 0  &        0 &     0    &   0    &  0   &   0         &     0         \\
              0  &   0   & P_{\bm{d}^{i}}^{\bm{d}^{i}}   &      0       &       0        &         0        &   0    &    0   &   0     &     0    &   0          &  0    &   0   &     0         \\
              0  &       0           &     P_{\bm{d}^{i}}^{\bm{d}^{if}}   & P_{\bm{d}^{if}}^{\bm{d}^{if}}  &  0     &         0        &         0          &     0    &    0     &     0    &   0          &     0    &      0      &     0         \\
              0  &       0           & 0  &  P_{\bm{d}^{if}}^{\bm{d}^f}    &  P_{\bm{d}^f}^{\bm{d}^f}  &         0        &         0          &     0    &    0     &     0    &   0          &     0    &      0      &     0         \\
\hline
\hline
  0  &   0   &           0   &   0              &   0            &    P_{\bm{u}^s}^{\bm{u}^s}  &   P_{\bm{u}^{is}}^{\bm{u}^s}     &   0  &    0     &     0    &   0          &     0    &      0      &     0         \\
   0  &   0 & 0   &   0              &   0            &    0  &   P_{\bm{u}^{is}}^{\bm{u}^{is}}  & P_{\bm{u}^{i}}^{\bm{u}^{is}}  &    0     &     0    &   0          &     0    &      0      &     0         \\
   0  &   0 & 0   &   0              &   0            &    0  &   0  & P_{\bm{u}^{i}}^{\bm{u}^{i}}   &    0     &     0    &   0          &     0    &      0      &     0         \\
   0             &       0           &   0   & 0  &    0   &         0        &         0          &    0   & P_{\bm{u}^{if}}^{\bm{u}^{if}}  &  0  &   0          &     0    &   0   &   0      \\
   0             &       0           & 0   &  0   &   0    &         0        &         0          &    0   &    P_{\bm{u}^{if} }^{\bm{u}^f }  &  P_{\bm{u}^f }^{\bm{u}^f }  &   0          &     0    &   0   &   0      \\
\hline
\hline
   0             &  0   &           0  &   0              &   0            &    0             &         0          &       0  &    0     &     0              &     P_{p^s}^{p^s} & P_{p^{is}}^{p^s}    &      0      &     0         \\
   0             &  0  &   0              &   0            &    0             &         0        &      0   &       0  &    0     &     0    &   0          &     P_{p^{is}}^{p^{is}} &      0      &     0         \\
   0             &       0           & 0   &       0       &         0       &      0 &      0  &         0          & 0  &    0     &   0          &     0    &      P_{p^{if}}^{p^{if}}      &     0      \\
   0             &       0           & 0   & 0 & 0   &         0       &         0          &   0  & 0   &  0 &   0          &     0    &     P_{p^{if}}^{p^{f}}      &       P_{p^{f}}^{p^{f}} \\
\end{bmatrix}
}
\,.
\end{equation}

The restriction operator is constructed in the usual way by evaluating coarse shape functions at fine nodes,
while the prolongator is the transpose of the restriction \cite{braess2007finite}.
We notice that certain blocks appearing in the restrictor are not present 
in the prolongator after transposition.
This is because certain nodes which are classified as  $is$ or $if$ at the coarse level
become $s$ or $f$ at the fine level, see Fig. \ref{domain_parts}.
 Also, notice that the two blocks 
 immediately to the right of $ R_{\bm{u}^{i}}^{\bm{u}^{i} } $ are zero, since coarse solid velocity equations are obtained 
  by restricting the fine solid velocity equations only, without taking contributions from the fluid part.
Finally, we remark that, when switching the blocks rows as in \eqref{jac_blocks_shuf},
the restriction and prolongation operators have to be changed accordingly.

\subsection{Richardson-Schwarz smoother}

\begin{figure}[htb]
\begin{center}
\includegraphics[scale=0.7]{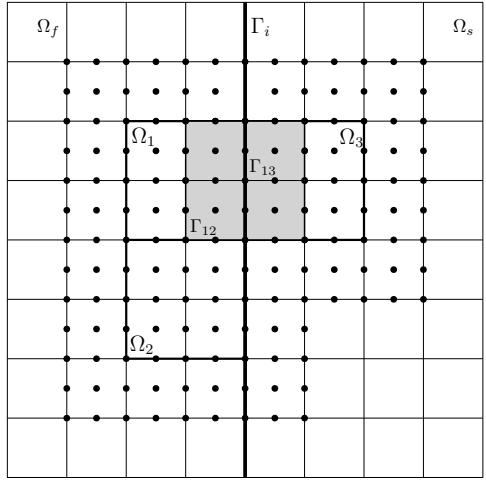} 
\caption{\review{Construction of the blocks for the domain decomposition in the additive Schwarz smoother:
\review{the domain $\Omega$ is split by $\Gamma_i$ into a solid part $\Omega_s$ and a fluid part $\Omega_f$ 
and then into smaller non-overlapping blocks;
$\Omega_1$, $\Omega_2$ and $\Omega_3$ are an example of admissible blocks (e.g. patches of four finite elements) 
with $\Gamma_{12}$ and $\Gamma_{13}$ being the boundaries between $\Omega_1$ and $\Omega_2$
and  between $\Omega_1$ and $\Omega_3$, respectively.
 The support of the test function associated to the midpoint of $\Gamma_{13}$
is highlighted in grey.}}}
\label{asm_dd}
\end{center}
\end{figure}

Here we define the smoother of the multigrid algorithm.
The coupled FSI system is treated in a monolithic manner at all stages,
except in the smoothing process.
We first partition the whole domain into the fluid and solid subregions, 
and then we further divide each subregion into 
smaller non-overlapping blocks ${\Omega}_k, k = 1, ...,N$, see Figure \ref{asm_dd}. 
The subdomain boundaries
are aligned with the triangulation, so that each ${\Omega}_k$ 
consists of an integral number of finite elements. 
%
 On each subdomain ${\Omega}_k$ 
we construct a subdomain preconditioner $\bm{B}_k$, 
which is a restriction of the Jacobian matrix $\bm{J}$, 
that is, it contains entries from $\bm{J}$ corresponding to the degrees of freedom 
contained in the corresponding subdomain ${\Omega}_k$. 
We define with $\bm{B}^f$ and $\bm{B}^s$ the restriction of the Jacobian matrix $\bm{J}$ to a fluid or solid subdomain.
The matrix block structures of  $\bm{B}^s$ and $\bm{B}^f$ as extracted from \eqref{jac_blocks_shuf} are given by
\begin{gather*}
\bm{B}^s =
\scalemath{0.74}{
\begin{bmatrix}[ ccc || ccc || cc]
   S_{\bm{d}^s}^{\bm{d}^s}     &   S_{\bm{d}^{is}}^{\bm{d}^s}    & 0                              &    S_{\bm{u}^s}^{\bm{d}^s}      & S_{\bm{u}^{is}}^{\bm{d}^s}       &      0                         &   S_{p^s}^{\bm{d}^s}         &      0                      \\
   S_{\bm{d}^s}^{\bm{d}^{is}}  &   S_{\bm{d}^{is}}^{\bm{d}^{is}} & S_{\bm{d}^{i}}^{\bm{d}^{is}}   &    S_{\bm{u}^s}^{\bm{d}^{is}}   & S_{\bm{u}^{is}}^{\bm{d}^{is}}    & S_{\bm{u}^{i}}^{\bm{d}^{is}}   &   S_{p^{s}}^{\bm{d}^{is}}    &  S_{p^{is}}^{\bm{d}^{is}}   \\
                            0  &   I_{\bm{d}^{is}}^{\bm{d}^{i}}  & I_{\bm{d}^{i}}^{\bm{d}^{i}}    &         0                       &   I_{\bm{u}^{is}}^{\bm{d}^{i}}   & I_{\bm{u}^{i}}^{\bm{d}^{i}}    &                   0          &  I_{p^{is}}^{\bm{d}^{i}}    \\
\hline
\hline
   K_{\bm{d}^s}^{\bm{u}^s}     &   K_{\bm{d}^{is}}^{\bm{u}^s}    &                            0   &        K_{\bm{u}^s}^{\bm{u}^s}    &   K_{\bm{u}^{is}}^{\bm{u}^s}     &   0                            &       0          &     0        \\
   K_{\bm{d}^s}^{\bm{u}^{is}}  &   K_{\bm{d}^{is}}^{\bm{u}^{is}} & K_{\bm{d}^{i}}^{\bm{u}^{is}}   &       K_{\bm{u}^s}^{\bm{u}^{is}}  &   K_{\bm{u}^{is}}^{\bm{u}^{is}}  & K_{\bm{u}^{i}}^{\bm{u}^{is}}   &      0          &     0         \\
                            0  &   K_{\bm{d}^{is}}^{\bm{u}^{i}}  & K_{\bm{d}^{i}}^{\bm{u}^{i}}    &                                0  &   K_{\bm{u}^{is}}^{\bm{u}^{i}}   & K_{\bm{u}^{i}}^{\bm{u}^{i}}    &      0          &     0         \\
\hline
\hline
   V_{\bm{d}^s}^{p^s} &  V_{\bm{d}^{is}}^{p^s} &           0           &    0             &         0          &       0  &    0     &     0                 \\
   0                  &  V_{\bm{d}^{is}}^{p^{is}} & V_{\bm{d}^{i}}^{p^{is}}  &   0                    &         0          &       0  &    0     &     0          \\
\end{bmatrix}
}
\quad
\bm{B}^f =
\scalemath{0.74}{
\begin{bmatrix}[ ccc || ccc || cc]
         I_{\bm{d}^{i}}^{\bm{d}^{i}}   & I_{\bm{d}^{if}}^{\bm{d}^{i}}  &       0        &   I_{\bm{u}^{i}}^{\bm{d}^{i}}   & I_{\bm{u}^{if}}^{\bm{d}^{i}}  &     0    &   I_{p^{if}}^{\bm{d}^{i}}   &     0         \\
               A_{\bm{d}^{i}}^{\bm{d}^{if}}   & A_{\bm{d}^{if}}^{\bm{d}^{if}}  &  A_{\bm{d}^f}^{\bm{d}^{if}}  &            0    &    0     &     0    &   0          &     0          \\
               0   & A_{\bm{d}^{if}}^{\bm{d}^f}  &  A_{\bm{d}^f}^{\bm{d}^f}  &            0    &    0     &     0    &   0          &     0        \\
\hline
\hline
    K_{\bm{d}^{i}}^{\bm{u}^{i}}   &   0              &   0            &    K_{\bm{u}^{i}}^{\bm{u}^{i}}   &    0     &     0    &   0          &     0       \\
    F_{\bm{d}^{i}}^{\bm{u}^{if}}   & F_{\bm{d}^{if}}^{\bm{u}^{if}}  & F_{\bm{d}^f}^{\bm{u}^{if}}   &        F_{\bm{u}^{i}}^{\bm{u}^{if}}   & F_{\bm{u}^{if}}^{\bm{u}^{if}}  &  F_{\bm{u}^f }^{\bm{u}^{if}}  &    F_{p^{if}}^{\bm{u}^{if}}   &   F_{p^f}^{\bm{u}^{if}}      \\
    0   & F_{\bm{d}^{if}}^{\bm{u}^f }  & F_{\bm{d}^f}^{\bm{u}^f }   &           0   & F_{\bm{u}^{if}}^{\bm{u}^f }  &  F_{\bm{u}^f }^{\bm{u}^f }  &    0   &   F_{p^f}^{\bm{u}^f }      \\
\hline
\hline
    W_{\bm{d}^{i}}^{p^{if}}  & W_{\bm{d}^{if}}^{p^{if}} &       0       &      W_{\bm{u}^{i}}^{p^{if}}  & W_{\bm{u}^{if}}^{p^{if}} &    0     &      0      &     0      \\
    0   & W_{\bm{d}^{if}}^{p^{f}} & W_{\bm{d}^f}^{p^{f}}  &          0  & W_{\bm{u}^{if}}^{p^{f}} &  W_{\bm{u}^f }^{p^{f}} &      0      &     0      \\
\end{bmatrix}
}
\,.
\end{gather*}

For the selected subdomain, the block structure of a submatrix $\bm{B}_k$ can assume one of the above two possibilities.   
In each block the equations to be solved are taken following a Vanka-type strategy.
The DOFs associated to an element consist of displacement, velocity and pressure.
While the support of the test functions associated to the pressure DOFs is limited to the element itself, 
the support of the test function associated to the displacement and velocity DOFs
extends to the neighboring elements. This extension is responsible for the overlapping and the exchange
of information between blocks. In Figure \ref{asm_dd} we highlighted in gray the support of the test function 
associated to the node in the middle of the edge $\Gamma_{13}$ separating $\Omega_1$ and $\Omega_3$.
Since only the weak mass balance equations are multiplied by the pressure test functions,
the communication between subdomains occurs only for the momentum balance and the kinematic equations.
 \review{A typical block size is obtained using 16 elements in 2D and 64 elements in 3D.
 According to the finite element discretization used in this work, these correspond to 372 and 4630 degrees of freedom per block,
 respectively}.
 %
 
 Given the preconditioned Richardson scheme 
for the Jacobian system $ \bm{J} \bm{e} = - \bm{r} $, 
\begin{displaymath}
 {\bm{e}}_{{j+1}}={\bm{e}}_{j} - \omega \bm{B}^{{-1}}( \bm{J} {\bm{e}}_{j} + {\bm{r}})  \, ,
\end{displaymath}
the restricted version of the additive Schwarz (ASM) preconditioner $\bm{B}^{-1}$ is
\begin{equation}
 \bm{B}^{-1} = \sum_{k=1}^N (\bm{R}^0_k)^T \bm{B}_k^{-1}(\bm{R}^{\delta}_k)\,.
\end{equation}

If $n$ is the total number of degrees of freedom in $\Omega$ and $n_k$ is the number of degrees of freedom in $\Omega_k$,
then $\bm{R}_k$ is a restriction matrix of size $ n_k \times n $ 
which maps the global vector of degrees of freedom to those belonging to the subdomain $\Omega_k$. 
Furthermore, $\bm{R}^{0}_k$ is a restriction matrix that does not include the overlap while $\bm{R}^{\delta}_k$ does. 
By looking at the matrix structure, $\bm{R}^{0}_k$ is the same as $\bm{R}^{\delta}_k$ with some rows set to zero.
The pattern above, that uses the overlap to provide information to the subdomain solver,
but that discards the result of that computation in the overlap region, 
is the restricted additive Schwarz algorithm \cite{barker2010}. 

Various inexact additive Schwarz preconditioners can be constructed 
by replacing the $\bm{B}_k$ matrices with ones that are convenient or inexpensive to compute.
From our numerical experiments we observed that the submatrices associated to fluid subdomains are \review{much less stiff} 
than the matrices associated to solid subdomains. 
While for the solid subproblems an LU decomposition is mandatory for the convergence of the overall algorithm,
for the fluid subproblems an ILU decomposition is sufficient to guarantee a good smoother behaviour.

\section{Numerical results}
\label{sec_numres}


\begin{figure}[htbp]
\centering
 \includegraphics[scale=0.40]{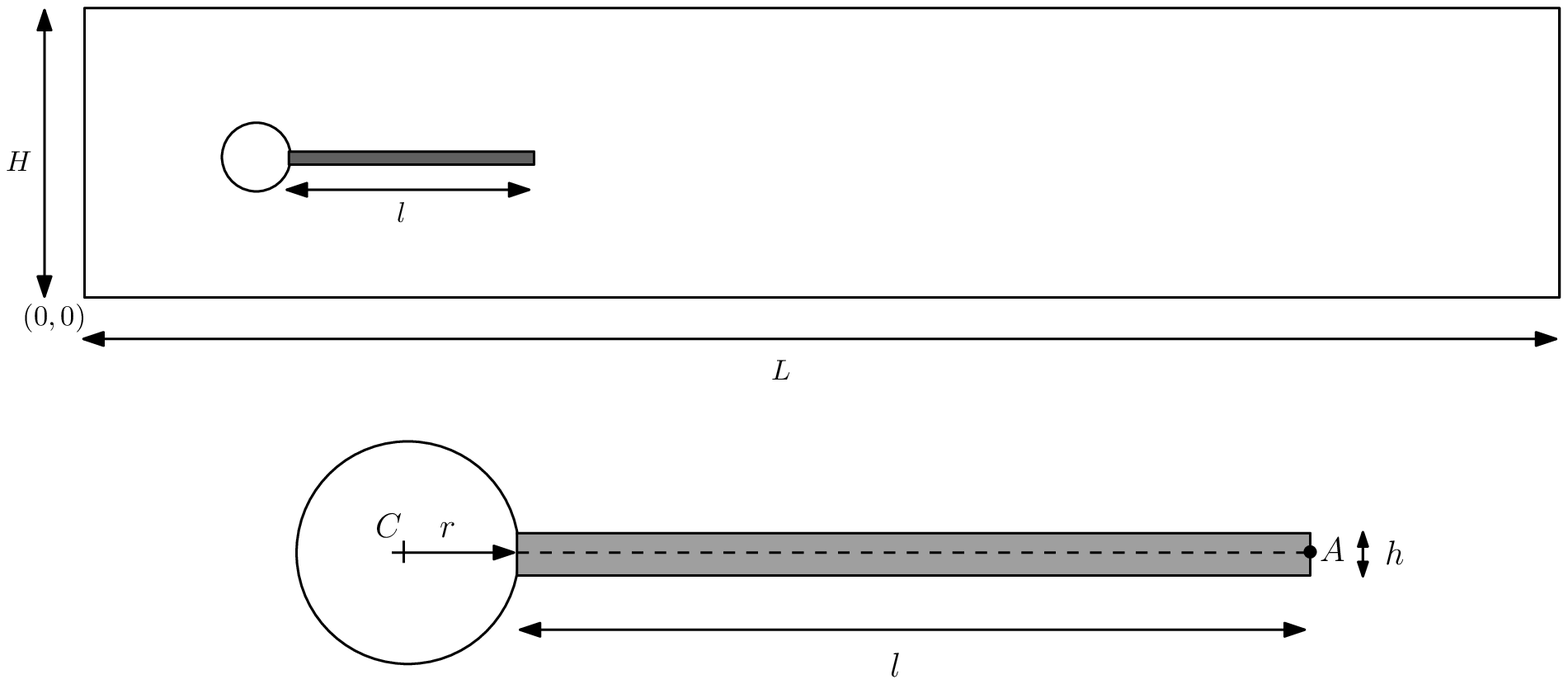} \\
 \includegraphics[scale=0.40]{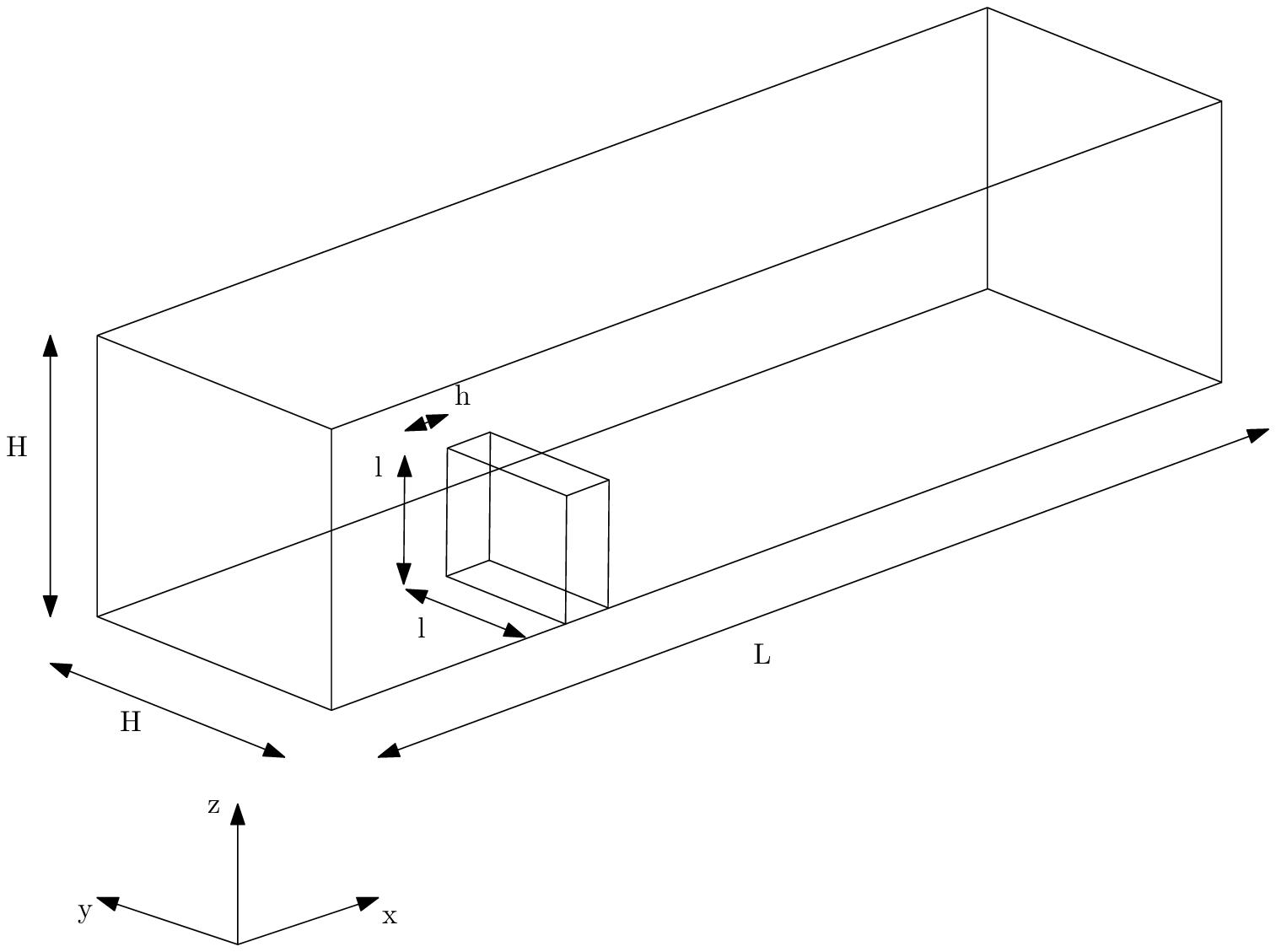} 
\caption{At the top: computational domain and detail of the beam for the 2D Hron-Turek FSI benchmarks. 
        At the bottom: computational domain for the 3D Richter FSI benchmarks.}
\label{figs:comp_domains}
\end{figure}

\begin{table}[!h]
\caption{\review{Number of unknowns of the 2D and 3D FSI benchmarks as a function of the number of mesh levels}}
\label{number_of_unknowns}
\begin{center}
\scalebox{0.7}{
\begin{tabular}{lrrrrrrr}
\toprule
Mesh level (2D)     & 1        & 2              & 3              & 4              & 5              & 6          \\ 
\midrule 
Unknowns            & 5 066    & 19 480         & 76 352         & 302 272        & 1 202 816      & 4 798 720  \\
\midrule
\midrule
Mesh level (3D)     & 1        & 2      & 3        & 4        & 5          &   \\ 
\midrule
Unknowns            & 2678     & 17 062 & 120 902  & 908 422  & 7 039 238  &   \\
\bottomrule
\end{tabular}
}
\end{center}
\end{table}
 
We seek to evaluate the accuracy, performance and robustness of the proposed Newton-Krylov solver
with Multigrid-Richardson-Schwarz preconditioner
for a set of FSI benchmark configurations that can be found in the literature, 
both for steady-state and for time-dependent cases,
both in two and three dimensions \cite{turek2006proposal, Richter}.
The sketches of the computational domains for the 2D and 3D cases 
are shown in Figure \ref{figs:comp_domains}. 
The two-dimensional benchmark geometry was proposed in \cite{turek2006proposal} for testing
and comparing different numerical methods and code implementations for fluid-structure interaction problems. 
The setup of this interaction problem consists of an elastic solid object attached
to a infinite rigid cylinder in a laminar channel flow. 
The three-dimensional benchmark consists of a flow past a deformable vertical wall 
and has been proposed and described by Richter in \cite{Richter}.

In 2D we consider three different test cases: a steady-state benchmark denoted as FSI1-S-2D and 
two time-dependent tests labelled FSI2-T-2D and FSI3-T-2D. 
The third one is a test case
that shows large deformations and stability problems caused by the added-mass effect \cite{Causin}.
In 3D we consider a stationary solution (FSI1-S-3D) 
and a nonstationary one (FSI2-T-3D).
The parameter settings of the benchmarks are collected
in Tables \ref{Tab:GeomDetails}, \ref{FSI_hronturek_paramsettings} and \ref{FSI_richter_paramsettings}.
In these tables, the nondimensional Young's modulus $ Ae $ is defined as $ Ae = \frac{E^s}{\rho^f U_m^2}$.
The number of unknowns associated to the mesh levels
for the Hron-Turek and the Richter benchmarks is collected in Table \ref{number_of_unknowns}.
All these tests were validated with respect to the available literature results. 
We refer to \cite{COMPDYN15FSI} for the validation of FSI1-S-2D.

\begin{table}[!h]
\caption{Geometrical parameters for the 2D and 3D cases.}
\label{Tab:GeomDetails}
\begin{center}
\scalebox{0.7}{
\begin{tabular}{lrrr}
\toprule
                              &  sym.  & 2D - value [m]    & 3D - value [m] \\ 
\midrule
channel length                &  L       &  $2.5$            & $1.5$  \\
channel width                 &  H       &  $0.41$           & $0.40$ \\
cylinder center position      &  C       &  $(0.2, 0.2)$     &  $-$   \\
cylinder radius               &  r       &  $0.05$           &  $-$   \\
elastic structure length      &  l       &  $0.35$           &  $0.2$ \\
elastic structure thickness   &  h       &  $0.02$           &  $0.1$ \\
reference point (at $t = 0$)  &  A       &  $(0.6, 0.2)$     &  $-$ \\
\bottomrule
\end{tabular}
}
\end{center}
\end{table}

Concerning the boundary conditions of the FSI problems in the steady-state cases,   
 $x$-velocity profiles $ u^{f}_{2D} $ and $ u^{f}_{3D} $ are prescribed in the inflow sections by
\begin{displaymath} 
    u^{f}_{2D}(0,y)   = \frac{3}{2} U_m \frac{y (H - y)}{\bigl(\frac{H}{2}\bigl)^2} \,, \quad
    u^{f}_{3D}(0,y,z) = \frac{9}{8} U_m \frac{z (H - z) (H^2 - y^2)}{\bigl(\frac{H}{2}\bigl)^2 H^2} \,.
\end{displaymath}
   In the time-dependent simulations both profiles are multiplied by $ \alpha ( t ) $
       given by 
\begin{equation}
    \alpha ( t ) = \begin{cases} 
                      \frac{1}{2} (1 - \cos (\pi t / 2 )) & \quad t < 2 s \\
                                   1                      & \quad t \geq 2 s \,,
                    \end{cases}
\end{equation}
   in order to smooth out the transition from $ {u} = {0} $ at $t = 0\,s$ to the maximum at $t = 2\,s$. 
  As an outflow boundary condition we set a stress-free boundary condition at the outflow section.
  The \textit{no-slip} condition is prescribed for the fluid on the other boundary edges in 2D:
  top and bottom wall, circle and fluid-structure interface. 
  In 3D the same boundary conditions are prescribed on all faces except for the face at $y = 0$
	where a symmetry condition is set.
We point out that the initialization of the nonlinear algorithm
is performed by prescribing zero velocity inside the fluid domain and no displacement in the structure.

\begin{table}[!h] 
\caption{Parameter settings for the Hron-Turek 2D FSI benchmarks.}
\label{FSI_hronturek_paramsettings}
\begin{center}
\scalebox{0.7}{
\begin{tabular}{lrrrrr}
\toprule
Parameter                   & sym.     & unit                 &  FSI1-S-2D                 &  FSI2-T-2D                &  FSI3-T-2D                 \\ 
\midrule
Fluid density               &  $\rho^f$  &  $[10^3 \frac{Kg}{m^3}]$     &   1                   &  1                   &   1                   \\
Fluid viscosity             &  $\nu^f$   &  $[10^{-3} \frac{m^2}{s}]$   &   1                   &  1                   &   1                   \\
\midrule
Solid density               &  $\rho^s$  &  $[10^3 \frac{Kg}{m^3}]$     &   1                   &  1                   &   1                   \\
Poisson coefficient         &  $\nu^s$   &  -                           &   0.5                 &  0.5                 &   0.5                 \\
Shear modulus               &  $\mu^s$   &  $[10^{6} \frac{Kg}{m s^2}]$ &   0.5                 &  0.5                 &   2.0                 \\
\midrule
Density ratio               &  $\beta$   &   -                          &   1                   &  1                   &   1                   \\
Dimensionless Young       &  $Ae$      &   -                          &   1.25 $\cdot 10^4$  &  5.0 $\cdot 10^2$   &   5.0 $\cdot 10^2$   \\
Avg. inlet velocity      &  $U_m$     &  $[\frac{m}{s}]$             &   0.2                 &  1                   &   2                   \\
Reynolds number             &  Re        &   -                          &   20                  &  100                 &   200                 \\
\bottomrule
\end{tabular}
}
\end{center}
\end{table}

\begin{table}[!h] 
\caption{Parameter settings for the Richter 3D FSI benchmarks.}
\label{FSI_richter_paramsettings}
\begin{center}
\scalebox{0.7}{
\begin{tabular}{lrrrr}
\toprule
Parameter                   & sym.     &  unit                 &  FSI1-S-3D                 &  FSI2-T-3D                 \\ 
\midrule
Fluid density               &  $\rho^f$  &  $[10^3 \frac{Kg}{m^3}]$     &   1                   &   1                   \\
Fluid viscosity             &  $\nu^f$   &  $[10^{-3} \frac{m^2}{s}]$   &   1                   &   1                   \\
\midrule
Solid density               &  $\rho^s$  &  $[10^3 \frac{Kg}{m^3}]$     &   1                   &   1                   \\
Poisson coefficient         &  $\nu^s$   &  -                           &   0.5                 &   0.5                 \\
Shear modulus               &  $\mu^s$   &  $[10^{6} \frac{Kg}{m s^2}]$ &   0.5                 &   0.5                 \\
\midrule
Density ratio               &  $\beta$   &   -                          &   1                   &   1                   \\
Dimensionless Young     &  Ae        &   -                          &   1.25 $\cdot 10^4$  &   5.0 $\cdot 10^2$   \\
Average inlet velocity      &  $U_m$     &  $[\frac{m}{s}]$             &   0.13333333          &   1                   \\
Reynolds number             &  Re        &   -                          &   26.67               &   100                 \\
\bottomrule
\end{tabular}
}
\end{center}
\end{table}

\subsection{Details on the iterative solver}

The nonlinear Newton convergence is detected at iteration $k \geq 1$ if
\begin{equation}
 || \bm{y}^{k} - \bm{y}^{k-1} ||_2 < \epsilon_{nl} \cdot || \bm{y}^{k} ||_2 \,,
\end{equation}
where the relative tolerance is $ \epsilon_{nl} = 10^{-8} $.
If we denote the action of the multigrid preconditioner with $ \bm{P} $, 
the linear convergence of the GMRES solver is achieved at iteration $j$ if
\begin{equation}
 || \bm{r}_L^j ||_2 < \epsilon_l \cdot || \bm{r}_L^{0} ||_2 \,,
\end{equation}
where $\epsilon_{l} = 10^{-4}$ is the relative linear tolerance and 
$\bm{r}_L^j = \bm{P}^{-1} (\bm{r}^j) $ is the left-preconditioned residual.
 Concerning the multigrid preconditioner, a direct solution is performed on the coarse mesh. 
 Four steps are taken of pre-smoothing and post-smoothing
 of a Richardson iteration preconditioned with ASM with a damping factor of $\omega = 0.7$.
 The size of each block is 16 elements in 2D and 64 elements in 3D
 and the local subproblem can be 
 solved exactly by a serial direct LU solver or approximately by an ILU(0) decomposition.

Our solver has been implemented in the open-source C++ Femus library (available at https://github.com/FeMTTU/femus), 
using the GMRES solver and the geometric multigrid preconditioner interface implemented in the PETSc toolkit \cite{petscuserref}.
The PETSc default options for GMRES were used 
(left preconditioner, classical unmodified Gram-Schmidt orthogonalization process) except for a lower restart (10). 
Some comments on parallelization are in order. 
PETSc offers a parallel implementation of the GMRES linear solver, so that the parallelization work for the user is left to the multigrid preconditioner part.
For the direct solver, we employed the implementation of the MUMPS package \cite{Mumps2001, Mumps2006} which includes a serial and a parallel version.
For the smoother, we adopt an ASM-preconditioned Richardson method for which the parallelization is based on the domain decomposition.
\review{In this paper only serial simulations were performed.}
We leave the testing and profiling of the parallel version of our solver to a future work. 

\begin{table}[!h]
\caption{Condition numbers of the linear system for all the simulations evaluated by the MUMPS package.}
\label{FSI_condnumber}
\begin{center}
\scalebox{0.7}{
\begin{tabular}{lrrrrr}
\toprule
Mesh level (FSI1-S-2D)     & 1                      & 2                      & 3                      & 4                     & 5                        \\ 
\midrule
average $ cond_1 $              & $9.21 \cdot 10^{5}$   & $3.70 \cdot 10^{6}$   & $1.60 \cdot 10^{7}$   & $5.99 \cdot 10^{7}$  & $2.42 \cdot 10^{8}$     \\                                               
max $ cond_1 $                  & $1.55 \cdot 10^{6}$   & $6.66 \cdot 10^{6}$   & $2.68 \cdot 10^{7}$   & $1.06 \cdot 10^{8}$  & $4.23 \cdot 10^{8}$     \\
average $ cond_2 $              & $2.99 \cdot 10^{5}$   & $1.52 \cdot 10^{6}$   & $1.54 \cdot 10^{7}$   & $9.02 \cdot 10^{8}$  & $6.69 \cdot 10^{9}$     \\                                               
max $ cond_2 $                  & $3.03 \cdot 10^{5}$   & $6.22 \cdot 10^{6}$   & $8.64 \cdot 10^{7}$   & $5.36 \cdot 10^{9}$  & $3.82 \cdot 10^{10}$     \\
\midrule
\midrule
Mesh level (FSI2-T-2D)     & 1                      & 2                      & 3                      & 4                     & 5                       \\ 
\midrule
average $ cond_1 $              & $-$                    & $-$                    & $3.85 \cdot 10^{4}$   & $1.53 \cdot 10^{5}$  & $4.63 \cdot 10^{5}$    \\
max $ cond_1 $                  & $-$                    & $-$                    & $2.61 \cdot 10^{5}$   & $1.05 \cdot 10^{6}$  & $4.23 \cdot 10^{5}$    \\
average $ cond_2 $              & $-$                    & $-$                    & $1.63 \cdot 10^{6}$   & $2.88 \cdot 10^{7}$  & $1.45 \cdot 10^{9}$    \\
max $ cond_2 $                  & $-$                    & $-$                    & $1.13 \cdot 10^{7}$   & $2.68 \cdot 10^{8}$  & $1.06 \cdot 10^{10}$    \\
\midrule
\midrule
Mesh level (FSI3-T-2D)     & 1                      & 2                      & 3                      & 4                     & 5                       \\ 
\midrule
average $ cond_1 $              & $-$                    & $-$                    & $6.62 \cdot 10^{4}$   & $2.82 \cdot 10^{5}$  & $9.98 \cdot 10^{5}$    \\
max $ cond_1 $                  & $-$                    & $-$                    & $4.37 \cdot 10^{5}$   & $1.62 \cdot 10^{6}$  & $6.53 \cdot 10^{6}$    \\
average $ cond_2 $              & $-$                    & $-$                    & $1.27 \cdot 10^{7}$   & $2.61 \cdot 10^{8}$  & $8.06 \cdot 10^{9}$    \\
max $ cond_2 $                  & $-$                    & $-$                    & $4.63 \cdot 10^{7}$   & $1.35 \cdot 10^{9}$  & $3.95 \cdot 10^{11}$    \\
\midrule
\midrule
Mesh level (FSI1-S-3D)     & 1                      & 2                      & 3                      &   &    \\ 
\midrule
average $ cond_1 $              & $8.04 \cdot 10^{1}$   & $7.44 \cdot 10^{2}$   & $3.12 \cdot 10^{3}$   &   &    \\
max $ cond_1 $                  & $4.57 \cdot 10^{2}$   & $4.28 \cdot 10^{3}$   & $1.78 \cdot 10^{4}$   &   &    \\
average $ cond_2 $              & $3.23 \cdot 10^{4}$   & $7.29 \cdot 10^{4}$   & $9.97 \cdot 10^{5}$   &   &    \\
max $ cond_2 $                  & $3.28 \cdot 10^{4}$   & $7.37 \cdot 10^{4}$   & $4.62 \cdot 10^{6}$   &   &    \\
\midrule
\midrule
Mesh level (FSI2-T-3D)     & 1                      & 2                      & 3                      &   &    \\ 
\midrule
average $ cond_1 $              & $-$                    & $1.23 \cdot 10^{2}$   & $2.76 \cdot 10^{2}$   &   &    \\
max $ cond_1 $                  & $-$                    & $2.56 \cdot 10^{2}$   & $1.29 \cdot 10^{3}$   &   &    \\
average $ cond_2 $              & $-$                    & $9.94 \cdot 10^{4}$   & $2.12 \cdot 10^{5}$   &   &    \\
max $ cond_2 $                  & $-$                    & $9.94 \cdot 10^{4}$   & $2.13 \cdot 10^{5}$   &   &    \\
\bottomrule
\end{tabular}
}
\end{center}
\end{table}

\subsection{Condition number estimates}
  
In order to assess the conditioning of the linear systems,
which influences the accuracy and convergence of the algorithms,
we computed estimates of condition numbers for all tests.
These estimates were obtained using two methods, 
 the \textit{condest} function provided by the MATLAB environment 
 and the error analysis performed by the MUMPS package,
 and are reported in Tables  \ref{FSI_condnumber_matlab} and \ref{FSI_condnumber}, respectively.
While the \textit{condest} function computes a lower bound 
for the condition number of the Jacobian matrix in the $l_1$-norm,
the error analysis performed by the MUMPS package
returns two condition numbers $ cond_1 $ and $ cond_2 $
which are related to the linear system and not just to the matrix. 
 We compared our results with \cite{Richter}. 
The condition numbers estimated by the MATLAB environment 
are in good agreement with the ones reported in \cite{Richter}.
However, the MUMPS values are smaller by some orders of magnitude than the Matlab ones
and are particularly small for the 3D tests.
This is a sign of good performance of both the direct and the iterative solver.

\begin{table}[!h]
\caption{Condition number estimate of the full system matrix $\bm{J}$ for all the simulations computed with Matlab. 
}
\label{FSI_condnumber_matlab}
\begin{center}
\scalebox{0.7}{
\begin{tabular}{lrrrrr}
\toprule
Mesh level (FSI1-S-2D)      & 1                      & 2                      & 3                      & 4                     & 5                       \\ 
\midrule
average $condest(\bm{J})$     & $1.41 \cdot 10^{12}$      & $1.10 \cdot 10^{13}$      & $9.85 \cdot 10^{13}  $       & $9.81 \cdot 10^{14}$   & $ 1.01 \cdot 10^{16} $   \\                                               
max $condest(\bm{J})$         & $1.43 \cdot 10^{12}$    & $1.21 \cdot 10^{13} $   & $1.13 \cdot 10^{14} $     & $1.11  \cdot 10^{15} $     & $ 1.14 \cdot 10^{16} $                     \\
\midrule
\midrule
Mesh level (FSI2-T-2D)      & 1                      & 2                      & 3                      & 4                     & 5                       \\ 
\midrule
average $condest(\bm{J})$     & $-$                    & $-$                    & $9.54  \cdot 10^{13}$     & $7.83  \cdot 10^{14} $                   & $-$                     \\
max $condest(\bm{J})$         & $-$                    & $-$                    & $9.68  \cdot 10^{13}$     & $8.42  \cdot 10^{14} $                   & $-$                     \\
\midrule
\midrule
Mesh level (FSI3-T-2D)     & 1                      & 2                      & 3                      & 4                     & 5                       \\ 
\midrule
average $condest(\bm{J})$     & $-$                    & $-$                    & $8.36 \cdot 10^{14}$         & $6.66 \cdot 10^{15}$                   & $-$                     \\
max $condest(\bm{J})$         & $-$                    & $-$                    & $8.43 \cdot 10^{14}$         & $6.73 \cdot 10^{15}$                   & $-$                     \\
\midrule
\midrule
Mesh level (FSI1-S-3D)     & 1                      & 2                      & 3                      &   &    \\ 
\midrule
average $condest(\bm{J})$     & $1.11 \cdot 10^{10}$  & $ 9.61 \cdot 10^{10} $   & $6.90  \cdot 10^{11} $         &   &    \\
max $condest(\bm{J})$         & $1.28 \cdot 10^{10}$  & $9.75  \cdot 10^{10} $   & $7.94  \cdot 10^{11} $         &   &    \\
\midrule
\midrule
Mesh level (FSI2-T-3D)    & 1                      & 2                      & 3                      &   &    \\ 
\midrule
average $condest(\bm{J})$     & $-$                    & $1.61 \cdot 10^{11}$                    & $1.16  \cdot 10^{12}$                    &   &    \\
max $condest(\bm{J})$         & $-$                    & $1.67 \cdot 10^{11}$                    & $1.18  \cdot 10^{12}$                    &   &    \\
\bottomrule
\end{tabular}
}
\end{center}
\end{table}

We remark that in these and in the following tables certain results are missing. 
This was due either to insufficient computational resources or to a too coarse mesh, 
inducing advection instabilities.
In the following tables, when we refer to average values for the steady-state cases 
we intend an average over the nonlinear iterations.
 For the time-dependent cases, the average is taken over 10 time steps based 
 on the average over the nonlinear iterations in each time step. 
 The time increments for the time-dependent tests are 
 $ \Delta t = 0.025 $ for FSI2-T-2D and  
  $ \Delta t = 0.01 $ for FSI3-T-2D and FSI2-T-3D.
  \review{The chosen timeframes are
   $[10.775,11.025]$  for FSI2-T-2D,
   $[6.42,6.52]$  for FSI3-T-2D and 
   $[3.6,3.7]$  for FSI2-T-3D.}
  For the steady-state cases, the number of nonlinear iterations
  is taken at the finest level of the Full Multigrid algorithm.
 For the time-dependent cases, the number of nonlinear iterations 
 is the most frequent number over the time steps. 
 We basically observed no variation between time steps.
   
The average convergence rate in the linear solvers in the following tables is defined as 
$
 \rho = \left(\dfrac{\bm{r}_N}{\bm{r}_0}\right)^{1/N}\,,
$
where $ N $ is the number of linear steps (in each nonlinear step).
 It is the geometric average of the ratios of residuals between two subsequent iterations.
 The average convergence rates for the direct solvers are very small 
 because the solution is obtained in one direct solver iteration,
 without additional iterative refinement steps.

\subsection{Comparison with a monolithic direct solver}

\begin{table}[!h]
\caption{Average convergence rate, memory usage, average computational 
time and Newton iterations for solutions using a monolithic direct solver.
}
\label{FSI_results_direct_solver}
\begin{center}
\scalebox{0.7}{
\begin{tabular}{lrrrrrrr}
\toprule
Mesh level (FSI1-S-2D)        & 1                     & 2                        & 3                       & 4                       & 5                             \\ 
\midrule
Average convergence rate       & $4.37 \cdot 10^{-11}$ & $2.54 \cdot 10^{-11}$    & $2.72 \cdot 10^{-10}$   & $2.31 \cdot 10^{-10}$   & $1.08 \cdot 10^{-09}$          \\
Solver memory usage  [MB]                 & 13                  & 57                     & 261                   & 1 187                 & 5 358                   \\
Average time  [s]                 & 0.05                 & 0.27                    & 1.45                   & 8.47                 & 54.74                          \\
Newton iterations               & 6                     & 6                        & 6                       & 6                       & 6                         \\
\midrule
\midrule
Mesh level (FSI2-T-2D)        & 1      & 2     & 3                      & 4                     & 5                               \\ 
\midrule
Average convergence rate       & $-$    & $-$   & $2.62 \cdot 10^{-12}$  & $3.95 \cdot 10^{-12}$ & $6.47 \cdot 10^{-12}$        \\
Solver memory usage  [MB]                  & $-$    & $-$   & 261                  & 1 187               & 5 358                 \\
Average time  [s]                 & $-$    & $-$   & 1.57                  & 9.38                 & 61.85                        \\
Newton iterations               & $-$    & $-$   & 5                      & 5                     & 5                           \\
\midrule
\midrule
Mesh level (FSI3-T-2D)        & 1      & 2     & 3                     & 4                       & 5                           \\ 
\midrule
Average convergence rate       & $-$    & $-$   & $3.45 \cdot 10^{-12}$ & $5.47 \cdot 10^{-12}$   & $1.07 \cdot 10^{-11}$        \\
Solver memory usage  [MB]                  & $-$    & $-$   & 261                 & 1 187                & 5 358             \\
Average time  [s]                 & $-$    & $-$   & 1.57                 & 9.45                   & 61.98                   \\
Newton iterations               & $-$    & $-$   & 5                     & 5                       & 5                       \\
\midrule
\midrule
Mesh level (FSI1-S-3D)        & 1                     & 2                     & 3                     & 4          & 5          \\ 
\midrule
Average convergence rate       & $2.19 \cdot 10^{-15}$ & $3.07 \cdot 10^{-14}$ & $5.78 \cdot 10^{-14}$ &  \review{$8.45 \cdot 10^{-14} $}       & $-$        \\
Solver memory usage  [MB]                  & 23                  & 257                 & 3 597               &  \review{42 285}   & $-$    \\
Average time [s]                  & 0.075                & 1.68                 & 86.86               &  \review{6877}       & $-$        \\   
Newton iterations               & 7                     & 7                     & 7                     &  \review{7}       & $-$        \\
\midrule
\midrule
Mesh level (FSI2-T-3D)        & 1      & 2                      & 3                       & 4           & 5          \\ 
\midrule
Average convergence rate       & $-$    & $2.19 \cdot 10^{-14}$  & $9.58 \cdot 10^{-14}$   &  $-$                     & $-$        \\
Solver memory usage  [MB]                  & $-$    & 257                  & 3 597                 &  $-$   &  $-$   \\
Average time  [s]                 & $-$    & 1.93              & 104.01                 &  $-$                        & $-$        \\
Newton iterations               & $-$    & 4                      & 4                       &  $-$       & $-$        \\
\bottomrule
\bottomrule
\end{tabular}
}
\end{center}
\end{table}

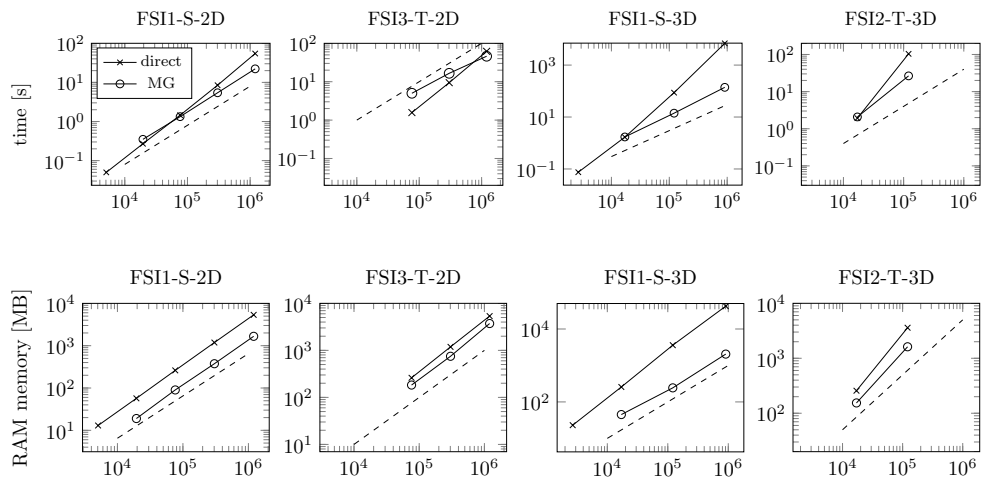
\begin{figure}[!h]
\captionsetup[subfigure]{labelformat=empty}
\centering
\begin{adjustbox}{max width=\textwidth}
\subfloat[][]{%
\begin{tikzpicture}
\begin{loglogaxis}[
xlabel={}, ylabel={time [s]},
title=FSI1-S-2D,  ymax=100, width=.37\textwidth,
legend style={at={(0.03, 0.97)}, anchor= north west, font=\footnotesize}
]
\addplot[black, mark=x, mark size=2.pt] coordinates {(5066, 0.05) (19480, 0.27) (76352, 1.45) (302272, 8.47) (1202816, 54.74)};
\addplot[black, mark=o, mark size=2.pt] coordinates {(19480, 0.35 ) (76352, 1.33) (302272, 5.43) (1202816, 22.21)};
\addplot[dashed ] coordinates {  (10000, 0.08) (100000, 0.8) (1000000, 8)};
\legend{direct, MG}
\end{loglogaxis}
\end{tikzpicture}
}
\subfloat[][]{%
\begin{tikzpicture}
\begin{loglogaxis}[
title=FSI3-T-2D,  xmin=3000, ymin=0.02,  ymax=100, width=.37\textwidth,
legend style={at={(0.03, 0.97)}, anchor= north west, font=\footnotesize}
]
\addplot[black, mark=x, mark size=2.5pt] coordinates {(76352, 1.57) (302272, 9.45) (1202816,  61.98)}; 
\addplot[black, mark=o, mark size=2.5pt] coordinates {(76352, 5.00) (302272, 16.56) (1202816,  46.10)};
\addplot[dashed ] coordinates {  (10000, 1) (100000, 10) (1000000, 100)};
\end{loglogaxis}
\end{tikzpicture}
}
\subfloat[][]{
\begin{tikzpicture}
\begin{loglogaxis}[
title=FSI1-S-3D,ymax=7000, width=.37\textwidth,
legend style={at={(0.03, 0.97)}, anchor= north west, font=\footnotesize}
]
\addplot[black, mark=x, mark size=2.pt] coordinates { (2678, 0.075) (17062, 1.68) (120902, 86.86) (908422, 6877)
 };
\addplot[black, mark=o, mark size=2.pt] coordinates {  (17062, 1.71) (120902, 14.06) (908422, 139) };
\addplot[dashed ] coordinates {  (10000, 0.3) (100000, 3) (1000000, 30)};
\end{loglogaxis}
\end{tikzpicture}
}
\subfloat[][]{%
\begin{tikzpicture}
\begin{loglogaxis}[
title=FSI2-T-3D,  ymin=0.03, xmin=2000,  ymax=200, width=.37\textwidth,
legend style={at={(0.03, 0.97)}, anchor= north west, font=\footnotesize}
]
\addplot[black, mark=x, mark size=2.pt] coordinates { (17062, 1.93) (120902, 104.01) };
\addplot[black, mark=o, mark size=2.pt] coordinates { (17062, 2.08) (120902, 26.53) };
\addplot[dashed ] coordinates {  (10000, 0.4) (100000, 4) (1000000, 40)};
\end{loglogaxis}
\end{tikzpicture}
}
\end{adjustbox}
\\
\begin{adjustbox}{max width=\textwidth}
\subfloat[][]{%
\begin{tikzpicture}
\begin{loglogaxis}[
xlabel={}, ylabel={RAM memory [MB]},
title=FSI1-S-2D,  ymax=10000, width=.37\textwidth,
legend style={at={(0.03, 0.97)}, anchor= north west, font=\footnotesize}
]
\addplot[black, mark=x, mark size=2.pt] coordinates { (5066, 13) (19480, 57) (76352, 261) (302272, 1187) (1202816, 5358) };
\addplot[black, mark=o, mark size=2.pt] coordinates { (19480, 19) (76352, 90) (302272, 375) (1202816, 1666) };
\addplot[dashed ] coordinates {  (10000, 6.5) (100000, 65) (1000000, 650)};
\end{loglogaxis}
\end{tikzpicture}
}
\subfloat[][]{
\begin{tikzpicture}
\begin{loglogaxis}[
title=FSI3-T-2D,  xmin=3000, ymin=7,  ymax=10000, width=.37\textwidth,
legend style={at={(0.03, 0.97)}, anchor= north west, font=\footnotesize}
]
\addplot[black, mark=x, mark size=2.pt] coordinates {(76352, 261) (302272, 1187) (1202816,  5358)};
\addplot[black, mark=o, mark size=2.pt] coordinates {(76352, 183) (302272, 749) (1202816,  3720)};
\addplot[dashed ] coordinates {  (10000, 10) (100000, 100) (1000000, 1000)};
\end{loglogaxis}
\end{tikzpicture}
}
\subfloat[][]{
\begin{tikzpicture}
\begin{loglogaxis}[
title=FSI1-S-3D, ymax=50000, width=.37\textwidth,
legend style={at={(0.03, 0.97)}, anchor= north west, font=\footnotesize}
]
\addplot[black, mark=x, mark size=2.pt] coordinates { (2678, 23) (17062, 257) (120902, 3597) (908422, 42285)};
\addplot[black, mark=o, mark size=2.pt] coordinates {  (17062, 45) (120902, 243) (908422, 2050)};
\addplot[dashed ] coordinates {  (10000, 10) (100000, 100) (1000000, 1000)};
\end{loglogaxis}
\end{tikzpicture}
} 
\subfloat[][]{%
\begin{tikzpicture}
\begin{loglogaxis}[
title=FSI2-T-3D,  xmin=1500, ymin=20, ymax=10000, width=.37\textwidth,
legend style={at={(0.03, 0.97)}, anchor= north west, font=\footnotesize}
]
\addplot[black, mark=x, mark size=2.pt] coordinates { (17062, 257) (120902, 3597) };
\addplot[black, mark=o, mark size=2.pt] coordinates { (17062, 154) (120902, 1619) };
\addplot[dashed ] coordinates {  (10000, 50) (100000, 500) (1000000, 5000)};
\end{loglogaxis}
\end{tikzpicture}
}
\end{adjustbox}
\caption{Computational time (top row) and memory usage (bottom) 
 as a function of the degrees of freedom
for the direct solver (x mark) and the multigrid-preconditioned GMRES solver (o mark) for four benchmark tests.
}
\label{FSI_direct_solver_plots}
\end{figure}

Here we describe the results of the FSI benchmark solutions. 
In order to assess the performance of the GMRES solver with Multigrid-Richardson-Schwarz preconditioner described in this paper, 
we compare it with a monolithic direct solver, implemented using the MUMPS package.
In Tables \ref{FSI_results_direct_solver} and \ref{FSI_results_mg} we collect the performance 
of the monolithic direct solver
and of the monolithic multigrid-preconditioned GMRES solver, respectively.
Computational time and memory usage are plotted in Figure \ref{FSI_direct_solver_plots}
for both solvers.
\review{
The computational time appears to have a linear behaviour for the iterative solver. 
The direct solver seems to be almost linear in time in 2D but with a superlinear behaviour in 3D.
The memory usage has a nearly linear behaviour for both the direct and the iterative solver.
 Also, we have a very good average convergence rate due to a low condition number of the linear system 
 (see Table \ref{FSI_condnumber}). }
This result is in contrast with what reported in \cite{Richter}.
Richter found out that with his formulation the high condition number of the matrix
causes a bad behaviour of the direct solver. 
All the efforts he made was 
to reduce the condition number of the linear problem by using the domain decomposition approach.
In our case, even if the matrix is ill-conditioned, the implementation of the MUMPS
direct solver was able to solve the linear problem without convergence issues.
This result opens the possibility to use a parallel direct solver for large problems 
by exploiting the high amount of memory and CPUs of modern supercomputers.

%



The observation of the results shows that the average convergence rates decrease with the mesh size.
Comparing the results of Table \ref{FSI_results_mg} 
with those of the direct solver in Table \ref{FSI_results_direct_solver}, 
we see a substantial improvement in memory consumption for all mesh sizes
but the improvement in terms of computational time is visible only for the finer meshes, in particular in 3D.
Comparing the total time with the average time we deduce that the cost of the matrix assembly using automatic differentiation
is high compared to the cost of the linear solution.

\begin{figure}[!t]
\begin{center}
\begin{adjustbox}{max width=0.7\textwidth}
\subfloat[][]{%
\begin{tikzpicture}
\begin{axis}[
title=, xlabel={time}, ylabel={}, ymax=50, legend style={at={(0.03, 0.97)}, anchor= north west, font=\footnotesize}
]
\addplot[black, mark=x, mark size=2.5pt] coordinates { (6.43, 29) (6.44, 27) (6.45, 26) (6.46, 31) (6.47, 38) (6.48, 44) (6.49, 45) (6.5, 42) (6.51, 40) (6.52, 34)};
\addplot[black, mark=o, mark size=2.5pt] coordinates {(6.43, 5) (6.44, 5) (6.45, 5) (6.46, 5) (6.47, 5) (6.48, 5) (6.49, 5) (6.5, 5) (6.51, 5) (6.52, 5)}; 
\legend{GMRES, Newton}
\end{axis}
\end{tikzpicture}
}\quad
\subfloat[][]{%
\begin{tikzpicture}
\begin{axis}[
title=, xlabel={time}, ylabel={}, ymax=20, legend style={at={(0.03, 0.97)}, anchor= north west, font=\footnotesize}
]
\addplot[black, mark=x, mark size=2.5pt] coordinates { (6.43, 16) (6.44, 16) (6.45, 15) (6.46, 15) (6.47, 16) (6.48, 16) (6.49, 14) (6.5, 18) (6.51, 17) (6.52, 16)};
\addplot[black, mark=o, mark size=2.5pt] coordinates {(6.43, 5) (6.44, 5) (6.45, 5) (6.46, 5) (6.47, 5) (6.48, 5) (6.49, 4) (6.5, 5) (6.51, 5) (6.52, 5)}; 
\legend{GMRES, Newton}
\end{axis}
\end{tikzpicture}
}
\end{adjustbox}
\end{center}
\caption{Number of Newton iterations and total number of GMRES linear iterations over 10 time steps
for the FSI3-T-2D case with 4 (a) and 5 (b) mesh levels. }
\label{FSI_results_mg_gmres}
\end{figure}
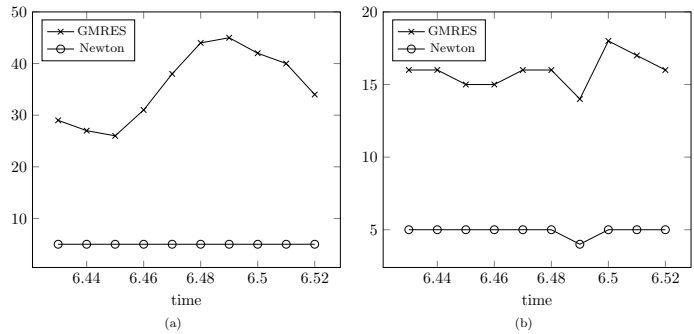

The number of nonlinear and linear iterations for 10 different time steps 
solved with the multigrid-preconditioned GMRES solver is shown 
in Figure \ref{FSI_results_mg_gmres} for two FSI3-T-2D tests.
The number of nonlinear iterations remains practically constant for all time steps and for all levels  
as we observe with the direct solver, 
while the number of GMRES linear steps varies with the time step 
due to a variation in time of the condition number of the matrix.

\subsection{Variation with respect to solver parameters}

\begin{table}[!h]
\caption{Characterization of the multigrid-preconditioned Newton-GMRES solver 
for various numbers of elements  $ N_{el,ASM} $ of each ASM block (FSI3-T-2D with 4 levels).}
\label{FSI_results_vanka_block}
\begin{center}
\scalebox{0.7}{
\begin{tabular}{lrrrr}
\toprule
 $N_{el,ASM}$                       & 4                     & 16                     & 64                     & 256                     \\ 
\midrule
Average convergence rate       & 0.32                  & 0.32                  & 0.32                  & 0.27                  \\
Solver memory usage [MB]            & 744                 & 749                 & 743                 & 930                 \\
Total memory usage  [MB]            & 2 048               & 1 817               & 1 819               & 2 006              \\
Average time [s]                  & 28.39                & 16.56               & 15.04              & 16.03                \\
Total time  [s]                    & 224.77               & 143.23               & 133.56               & 138.08               \\
Newton iterations              & 5                     & 5                     & 5                     & 5                     \\ 
\bottomrule
\end{tabular}
}
\end{center}
\end{table}

\begin{table}[!h]
\caption{Characterization of the multigrid-preconditioned Newton-GMRES solver for various numbers
 of pre-smoothing and post-smoothing steps $ (N_{pre},N_{post}) $  (FSI3-T-2D with 4 levels).
}
\label{FSI_results_pre_post}
\begin{center}
\scalebox{0.7}{
\begin{tabular}{lrrrrr}
\toprule
 $ (N_{pre},N_{post}) $          & (2,2)                   & (2,4)                   & (4,2)                   & (4,4)        & (8,8)                 \\ 
\midrule
Average convergence rate       & 0.61                  & 0.47                  & 0.46                  & 0.32       & 0.19                \\
Average linear steps           & 18                    & 12                    & 11                    & 7          & 5                   \\
Solver memory usage [MB]           & 797                 & 783                 & 782                 & 776     & 753              \\
Total memory usage [MB]             & 1 873              & 1 859              & 1 858              & 1 852    & 1 830            \\
Average time [s]                  & 17.63                & 18.23               & 17.95               & 16.78    & 23.85             \\
Total time   [s]                    & 149.26              & 156.62              & 152.82             & 145.55   & 176.87          \\
Newton iterations              & 5                     & 5                     & 5                     & 5          & 5                   \\ 
\bottomrule
\end{tabular}
}
\end{center}
\end{table}

In order to study the performance of the iterative solver with respect to some solver parameters,
\review{Tables} \ref{FSI_results_vanka_block} and \ref{FSI_results_pre_post} 
show the behaviour 
using different numbers of elements $N_{el,ASM}$ of each ASM subdomain  
and different numbers of pre-smoothing  $ N_{pre} $ and post-smoothing  $ N_{post} $ steps.
Each ASM subdomain has the same number of elements $ N_{el,ASM} $ given by 
\begin{equation}
 N_{el,ASM} = (2^n)^{\theta_{ASM}} \,,
\end{equation}
where $n=2$ or $ n=3 $ is the space dimension and $ \theta_{ASM} $ is a variable exponent.
Table \ref{FSI_results_vanka_block} shows that the average convergence rate
is not much influenced by the size of the subdomain.
The solver average time is highest for the smallest ASM subdomain 
and has a minimum for a subdomain made of $64$ elements,
as already found in \cite{SimonePhD}.
This size is also optimal from the memory consumption point of view. 
We believe that this corresponds to
an optimal condition for minimizing the overall time 
of extraction of the submatrices
and subsequent LU decomposition.
Table \ref{FSI_results_pre_post} shows instead that both the convergence rate and the average time are influenced by the number of 
pre-smoothing and post-smoothing steps. On the contrary the memory consumption remains almost constant for all values.
Even if the convergence rate decreases with the number of pre-smoothing and post-smoothing steps, 
a minimum of the average solver time is reached with the combination of four pre- and post-smoothing steps.

\subsection{Comparison with literature numerical studies}

\begin{table}[!h]
\caption{Average convergence rate, memory usage, average computational 
time and Newton iterations for solution with our multigrid-preconditioned Newton-GMRES solver  
($\theta_{ASM} = 2$, $ N_{pre} = N_{post} = 4 $)
}
\label{FSI_results_mg}
\begin{center}
\scalebox{0.7}{
\begin{tabular}{lrrrrr}
\toprule
Mesh level (FSI1-S-2D)        & 2                     & 3                     & 4                     & 5               \\ 
\midrule
Average convergence rate       & 0.15                  & 0.1                   & 0.09                  & 0.08            \\
Solver memory usage  [MB]            & 19                  & 90                    & 375                   & 1 666          \\
Total memory usage   [MB]            & 105                   & 376                   & 1450                  & 5 713    \\
Average time  [s]                  & 0.35                 & 1.33                 & 5.43                 & 22.21         \\
Total time    [s]                  & 5.58                 & 23.09                & 77.69                & 314.75        \\
Newton iterations            & 4                     & 4                     & 3                     & 3                \\
\midrule
\midrule
Mesh level (FSI2-T-2D)        & 2                     & 3                     & 4                     & 5                \\ 
\midrule
Average convergence rate        & $-$                   & 0.08                  & 0.05                  & 0.04           \\
Solver memory usage   [MB]      & $-$                   & 178                   & 753                   & 3 470          \\
Total memory usage    [MB]      & $-$                   & 453                   & 1 820                 & 7 205           \\
Average time   [s]              & $-$                   & 2.62                  & 10.52                 & 45.16           \\
Total time    [s]               & $-$                   & 28                   & 113                  & 601               \\
Newton iterations               & $-$                   & 5                     & 5                     & 6              \\       
\midrule
\midrule
Mesh level (FSI3-T-2D)        & 2                     & 3                     & 4                     & 5                \\ 
\midrule
Average convergence rate       & $-$                   & 0.46                  & 0.33                  & 0.07            \\
Solver memory usage   [MB]           & $-$                   & 183                   & 749                   & 3 720    \\
Total memory usage    [MB]           & $-$                   & 458                   & 1 817                 & 7 276    \\
Average time   [s]                   & $-$                   & 5.00                  & 16.56                 & 46.10     \\
Total time    [s]                    & $-$                   & 39.60                 & 143.23                & 492.88    \\
Newton iterations               & $-$                   & 5                     & 5                     & 5              \\       
\midrule
\midrule
Mesh level (FSI1-S-3D)        & 2                     & 3                     & 4           & 5          \\ 
\midrule
Average convergence rate       & 0.25                  & 0.22                  &  \review{0.25}        & $-$        \\
Solver memory usage   [MB]           & 45                    & 243                   &  \review{2050}      & $-$      \\
Total memory usage    [MB]           & 325                   & 2275                  &  \review{17858}     & $-$      \\
Average time   [s]                 & 1.71                 & 14.06                &  \review{139}        & $-$        \\
Total time     [s]                 & 73.60                & 711.61              &  \review{3848}        & $-$        \\ 
Newton iterations             & 5                     & 5                     &  \review{5}        & $-$        \\
\midrule
\midrule
Mesh level (FSI2-T-3D)        & 2                     & 3                     & 4           & 5          \\ 
\midrule
Average convergence rate       & 0.06                  & 0.08                  &  $-$        & $-$        \\
Solver memory usage   [MB]           & 154                   & 1 619                 &  $-$      & $-$      \\
Total memory usage    [MB]           & 425                   & 3 641                 &  $-$      & $-$      \\
Average time   [s]                 & 2.08                 & 26.53                &  $-$        & $-$        \\
Total time    [s]                  & 50.29               & 494.62               &  $-$        & $-$        \\
Newton iterations           & 4                     & 4                     &  $-$        & $-$        \\
\bottomrule
\bottomrule
\end{tabular}
}
\end{center}
\end{table}



Let us compare our work with other numerical studies in the literature, such as \cite{gee2011truly}, \cite{Richter} and \cite{Muddle12}.
Our setting of the 2D and 3D cases is the same as \cite{Richter}, in order to establish an appropriate comparison. 
The two-dimensional example discussed by Gee, K\"uttler and Wall in section 8.2 of \cite{gee2011truly}
is comparable with our 2D benchmark problems.
 In all the configurations, a self-excited motion of a flexible beam in a laminar flow is described.
In \cite{gee2011truly} a discretization with 80000 degrees of freedom is used, 
which corresponds to mesh level 4 in \cite{Richter} (Table II) 
and to mesh level 3 in our case (Table \ref{number_of_unknowns}).
On this mesh, the solver proposed in our paper requires an average of 5 s.
An average of five Newton steps per time-step results in 25 s per time step 
compared with an average of 9.5 s taken from \cite{Richter} and 7 s taken from \cite{gee2011truly}.
Our solver reveals to be around 3 times slower than the results found by \cite{Richter} and \cite{gee2011truly}.
A direct comparison is however not an easy task: \cite{gee2011truly} used the same Newton tolerance of $10^{-4}$ 
but a parallel solver on four cores, while our results are reported in single core performance. 
Assuming a linear scaling,
their single core performance is in agreement with our result. 
The performance of the Geometric Multigrid proposed in \cite{Richter} should be appreciated since the author 
uses a single core and a more restrictive Newton tolerance.
Let us also remember that our solver enforces full incompressibility both for the fluid and for the solid
without stabilization terms.

In \cite{Muddle12} (Section 4.3.1), the authors investigate 
a partitioned scheme as preconditioner for a monolithic GMRES iteration.
Comparable with our strategy and \cite{Richter}, 
the authors report the measures of the performance of their solver which uses a direct inversion of the subsystems.
They show nearly robust and good convergence properties. 
However, they find that the convergence rate does not improve with finer meshes. 
This result is in contrast to what \cite{Richter} and us have reported in Table V and \ref{FSI_results_mg}, respectively.
Thus, our results confirm the remark reported in \cite{Richter} (remark 3 section 5.2) 
that a preconditioner to a monolithic GMRES iteration composed by a multigrid solver 
with a partitioned smoother (in our case with exact solution of only the solid subproblems) 
is more robust than one based on a partitioned solver.

\section{Conclusions}

In this paper we studied a monolithic Newton-Krylov solver
 preconditioned by geometric multigrid for incompressible FSI problems.
 Modified Richardson smoothers preconditioned by an additive Schwarz algorithm were considered. 
Subdomain blocks are extracted from the exact Jacobian matrix
in order to define the Schwarz algorithm in the smoothing process.
An extensive analysis
with condition number estimates, time performance, RAM memory usage and variation of solver parameters
has been carried out on several two- and three-dimensional FSI benchmark tests.
A very good behaviour of our solver is observed
both for steady-state and time-dependent tests.
In particular, we emphasize that the results were obtained 
with a fully incompressible formulation without stabilization terms 
and also for direct-to-steady-state simulations.
Further lines of investigations such as the adoption of different orderings of the degrees of freedom,
adaptive refinement strategies
and multigrid advection stabilization algorithms are under current inspection.

\bibliographystyle{plain}
\bibliography{fsi}

\end{document}